\documentclass[amsmath,amssymb,jcp,aip,twocolumn,mathbbm, reprint, superscriptaddress]{revtex4-2}

\usepackage{amsmath,amssymb}
\usepackage{bbm}
\usepackage{fontspec}
\usepackage{graphicx}
\usepackage{booktabs}
\usepackage{csquotes}
\usepackage{xcolor}
\usepackage{mathtools}
\usepackage{physics}
\usepackage{siunitx}
\usepackage{chemformula}
\usepackage{listings}
\usepackage[inline]{enumitem}
\usepackage{hyperref}

\newfontfamily\ubuntumono{DejaVuSansMono.ttf}
\definecolor{codegreen}{rgb}{0,0.6,0}
\definecolor{codegray}{rgb}{0.5,0.5,0.5}
\definecolor{codepurple}{rgb}{0.58,0,0.82}
\definecolor{backcolour}{rgb}{0.95,0.95,0.92}

\hypersetup{
    colorlinks,
    linkcolor={red!50!black},
    citecolor={blue!50!black},
    urlcolor={blue!80!black}
}

\newcommand{\fleft}{\mathopen{}\mathclose\bgroup\left}
\newcommand{\fright}{\aftergroup\egroup\right}

\graphicspath{{./figs/}}

\DeclareMathOperator{\spn}{span}

\newcommand{\mname}{\textit{paces}}
\newcommand{\qtrue}{$q_\text{true}$}
\newcommand{\qnom}{$q_\text{nom}$}

\makeatletter
\def\@email#1#2{%
 \endgroup
 \patchcmd{\titleblock@produce}
  {\frontmatter@RRAPformat}
  {\frontmatter@RRAPformat{\produce@RRAP{*#1\href{mailto:#2}{#2}}}\frontmatter@RRAPformat}
  {}{}
}%
\makeatother

\begin{document}

\title{\textit{paces}: Parallelized Application of Co-Evolving Subspaces, a method for computing quantum dynamics on GPUs} 

\author{R.~Kevin~Kessing}
\email{kevin.kessing@uni-ulm.de}
\affiliation{Institut f\"ur Theoretische Physik, Universit\"at Ulm, 89069 Ulm, Germany} 
\affiliation{Institut f\"{u}r Theoretische Physik, Georg-August-Universit\"{a}t G\"{o}ttingen, 37077 G\"{o}ttingen, Germany}
\affiliation{Department of Chemistry, Massachusetts Institute of Technology, Cambridge, Massachusetts 02139}

\begin{abstract}
An efficient method of solving the time-dependent Schrödinger equation for pure states is described:
At each timestep, a restricted subspace of the total Hilbert space is systematically and naturally
constructed via the image of repeated applications of the Hamiltonian operator,
and the time evolution is computed exactly within said subspace.
The subspace is dynamically recomputed such that it co-evolves with the state vector.
The method is built from the ground up as a parallel algorithm for graphics processing units
and suited to Hamiltonians that are sparse in a given basis.
We benchmark the method by comparing its results for a 1D Holstein model to previously published multiset-MPS results
and then apply the method to compute optical spectra and non-equilibrium dynamics of one-, two- and three-dimensional
model chromophore nanoaggregates.
\end{abstract}

\maketitle

\begin{figure*}
\centering
\includegraphics{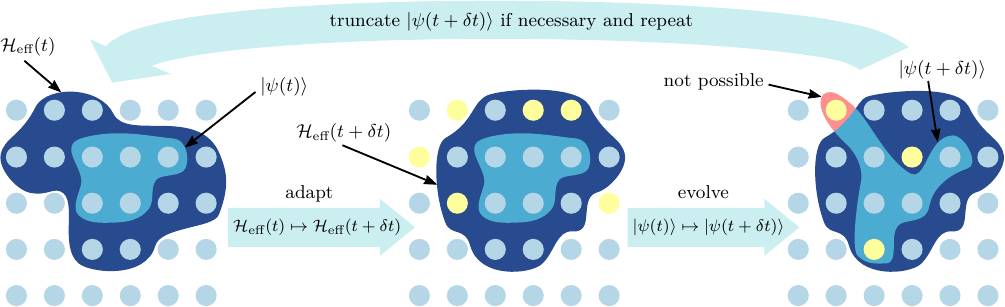}
\caption{The basic idea of \mname{}: Construct an adapted effective Hilbert space $\mathcal{H}_\text{eff}$,
evolve exactly within this effective Hilbert space, truncate and repeat. Each circle represents a basis state.
Dark blue areas: the effective Hilbert subspace;
light blue areas: non-zero components of the state vector;
yellow highlighting: basis states affected by a change.
The red area lies outside of $\mathcal{H}_\text{eff}$ and cannot be occupied by the state vector at this stage.}
\label{fig:basic_idea}
\end{figure*}
\section{Introduction \& motivation}\label{sec:intro}
The dynamics of any closed quantum system are described by the time-dependent Schrödinger equation (TDSE),
\[i\hbar \pdv{t} \ket{\psi(t)} = H \ket{\psi(t)},\]
whose solution is formally---if the Hamiltonian operator $H$ is time-independent---given by
\begin{equation}
\ket{\psi(t)} = e^{-i t H/\hbar} \ket{\psi(0)}, \label{eq:time_evolution}
\end{equation}
where $\ket{\psi(0)}$ is the initial state that is to be evolved in time
and $\ket{\psi(t)}$ is the state after evolving for some time~$t$.
In practice, equation~\eqref{eq:time_evolution} usually cannot be computed exactly
due to the well-known \enquote{curse of (exponential) dimensionality}~\cite{OseledetsDim,Wang2015,Borrelli2016,Xie2019},
i.e., the exponential growth of the Hilbert space with respect to the system size,
which prohibits a complete analysis of all but the simplest problems (e.g., a laser-driven \ch{H2} molecule~\cite{H2sim}).
Therefore, a number of approaches have been developed that effectively try to reduce the size of the problem to a manageable scale:
Important examples of such methods are matrix-product state (MPS) approaches~\cite{Bridgeman,Schollwoeck,Paeckel2019}
which employ local decompositions and restrictions to lower-entanglement states,
or the multi-configuration time-dependent Hartree (MCTDH) method~\cite{Meyer1990CPL,Manthe1992JCP,Meyer2009}
as well as a variety of methods based on Gaussian wavepackets~\cite{Shalashilin2000,Shalashilin2006,Ben-Nun,Gu2016,Werther2020,Pettey2006,Pettey2007,Wu2003,Wu2004,Chen2006,Koch2013,Saller2017,Richings},
which, in a sense, increase the efficacy of a small basis set by allowing the basis functions themselves to adapt.
Moreover, other methods such as sparse polynomial methods~\cite{Alvermann2008PRB,Alvermann2009PRL}
instead treat a large number of degrees of freedom using efficient sparse interpolations.

In this article, we present a highly parallelized method of solving the TDSE based on adaptive or \emph{co-evolving} subspaces
developed for use on graphics processing units (GPUs).
This method has been successfully applied in a previous publication~\cite{Kessing2022JPCL},
but the present publication constitutes the first detailed explanation and investigation of its strengths and weaknesses.
The Python/CuPy~\cite{cupy}-based code that was used to create all numerical data shown in this publication is
publicly available in the \href{https://github.com/rkevk/paces}{\mname{} repository} on GitHub.

In contrast to MPS methods, the present method performs no local decompositions and uses a much simpler compression scheme,
and it is distinct from most MCTDH and wavepacket-type approaches in that it employs time-independent basis states,
simply adding and removing basis states as required without changing the states themselves
(though a wavepacket-based method known as \textsc{proDG}~\cite{Hartke2006,Sielk2009} notably also does so).
Systematic grow-and-truncate approaches have also been applied in other contexts, such as the dynamic filtering approach
in hierarchical equations of motion~\cite{Chen2009JCP,Shi2009JCP}, corner-space renormalization~\cite{Finazzi2015PRL}
and the highly efficient kernel polynomial method~\cite{Weisse2006RMP}, but unlike these methods,
the present method directly computes a time-evolved pure quantum state.
As such, the present method bears resemblance to
recent time-dependent adaptive (sampling) configuration interaction approaches~\cite{Schriber2019JCP,Shee2025JCP},
but it is set apart from these by the recurring adaptation and implementation details
such as the nature of the time-evolution step and the focus on parallelization.

A particular method worth highlighting in the context of our method is the limited functional space (LFS) method
developed by Bon\v{c}a et al.~\cite{Bonca} and applied in many subsequent publications, e.g.,~\onlinecite{%
Bonca2000PRL,Ku2002,Bonca2008PRB,Vidmar2009PRL,Vidmar2011PRB,Vidmar2011PRL,Golez2012PRL,Vidmar2013,FHM2015}.
The core idea of LFS methods is also based on restricted Hilbert spaces that are determined
via repeated application of the system Hamiltonian.
Furthermore, a later variant of LFS called self-consistent variational exact diagonalization~\cite{Chakraborty2013PRB,Chakraborty2014PRB,Chakraborty2016PRB}
used this principle variationally and dynamically for exact diagonalization,
similar in spirit, as we will describe below, to our repeated application of the Hamiltonian---%
though the present method focuses on parallelized calculations of time evolution.

This paper is structured as follows:
In section~\ref{sec:the_essence}, we provide a high-level overview of the idea of the method
which is then explained in extensive detail in section~\ref{sec:details}.
The latter is split in three parts:
The first, section~\ref{sec:info_comparison}, motivates the use of a vector-based representation
as employed by the current method and discusses when or when not it might prove to be useful.
Section~\ref{sec:sparseHam} briefly discusses some aspects of sparseness in many-body systems and the related concept
of what we term connectivity, setting the stage for section~\ref{sec:adaptation},
which introduces the concepts of co-evolving subspaces and shows how they are applied in the present method.
Section~\ref{sec:performance} quantitatively estimates errors incurred by this method and then benchmarks against
existing multiset-MPS calculations.
In section~\ref{sec:aggregates}, we apply the method to compute optical spectra and non-equilibrium dynamics
of model vibronic nanoaggregates in one, two and three dimensions,
before concluding by discussing potential future applications of the method in section~\ref{sec:conclusion}.

\section{The essence of \mname{}: Following a wavefunction as it propagates}\label{sec:the_essence}
The basic idea of the present method is to efficiently truncate a high-dimensional underlying Hilbert space
to a smaller Hilbert space in a time-dependent manner.
This is similar to the rationale of MPS methods~\cite{Manmana2005,Paeckel2019};
however, unlike the SVD-based compression of MPS methods,
the present method uses a different expansion-and-truncation technique to manage the effective Hilbert space.

The core principles can be described as follows (illustrated in Fig.~\ref{fig:basic_idea}):
In a finite-dimensional Hilbert space $\mathcal{H}$ spanned by a countable orthonormal basis $\mathfrak{B}$,
assume the initial state vector $\ket{\psi}$ has non-zero basis coefficients $\psi_n = \braket{n}{\psi}$ for only a small number of basis states $\ket{n} \in \mathfrak{B}$. Then:
\begin{enumerate}
\item \label{special:step1} \emph{construct a Hilbert subspace}, called the \emph{effective Hilbert space} $\mathcal{H}_\text{eff}(t)$, of the entire $\mathcal{H}$; this Hilbert subspace should capture both the current state vector and \enquote{neighboring} basis states into which the state vector may evolve,
\item then construct the Hamiltonian matrix acting on this Hilbert subspace and \emph{evolve} the state vector based on the Hamiltonian within the subspace (using $U(\delta t) = e^{-i H \delta t /\hbar}$),
\item and finally, \emph{truncate} the resulting state vector so as to prevent a memory overflow in the following iteration,
then repeat from step~\ref{special:step1}.
\end{enumerate}
The notion of a \enquote{neighboring} basis state is defined generically in terms of the image of the Hamiltonian
acting on the state at time $t$, as explained below.

Note that all three steps are repeated at every timestep,
such that the effective subspace is repeatedly reconstructed in a co-evolving manner.
This, together with its emphasis on parallelized deployment on GPUs, lends it the designation
\enquote{\textbf{pa}rallelized \textbf{c}o-\textbf{e}volving \textbf{s}ubspaces} or \mname{}.

The motivation for the recurring and adaptive recalculation of the relevant subspace is that most of a Hilbert space is
irrelevant to finite-time dynamics~\cite{Poulin2011PRL,Verstraete}:
Given an initial state in an exponentially large Hilbert space and a realistically scaled Hamiltonian,
the time it would take to evolve into most parts of the Hilbert space would far exceed the relevant timescales,
such that only an exponentially tiny fraction of the Hilbert space is relevant for finite-time evolution.
\begin{figure}
\centering
\includegraphics{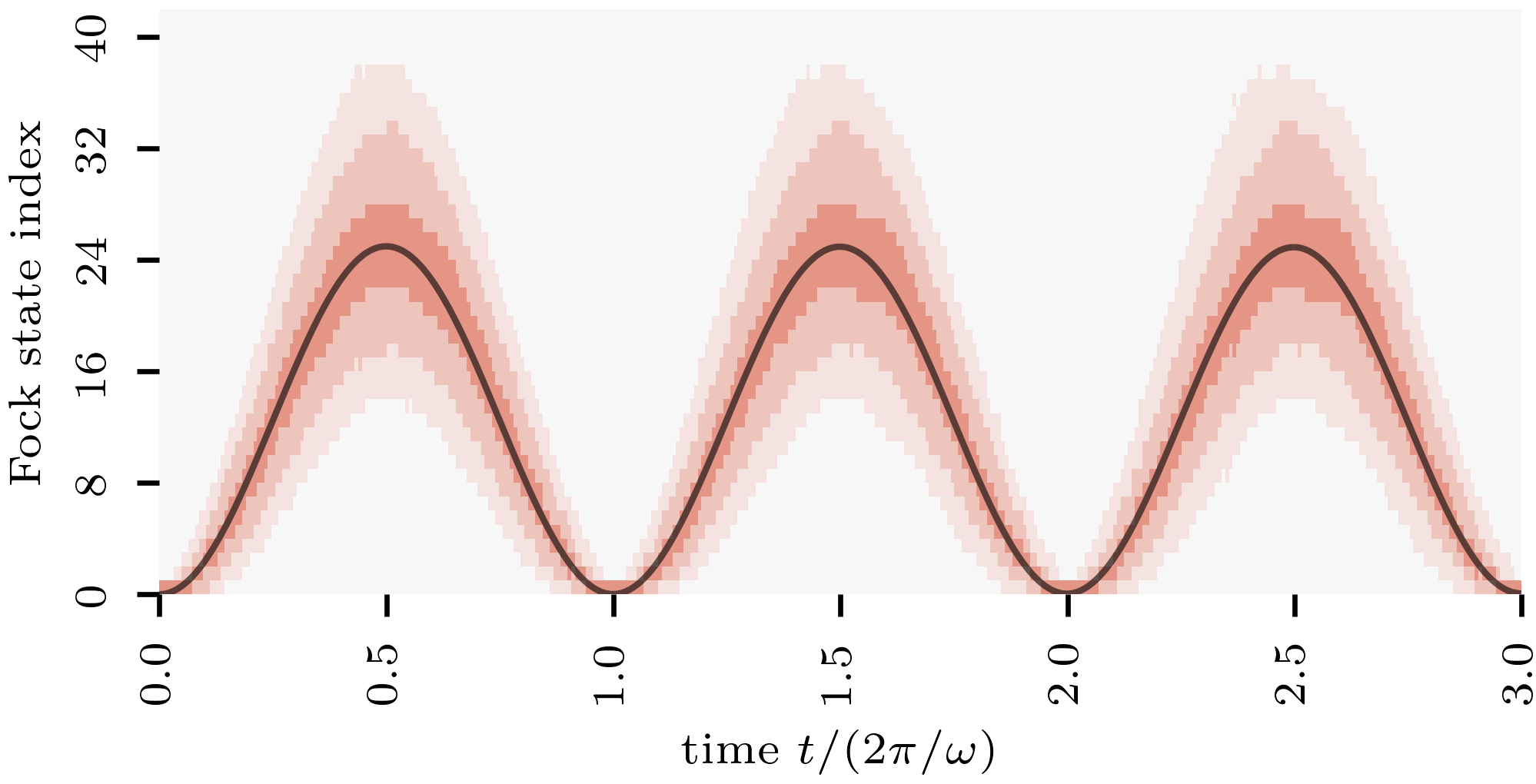}
\caption{The physical foundation of \mname{}---i.e., adaptive restricted basis sets---%
illustrated using a shifted quantum harmonic oscillator
$H = \hbar\omega \left( a^\dag a + 2.5\left(a + a^\dag\right)\right)$ starting from an initial vacuum state $\ket{0}$.
The dark line is the average occupation number $\expval{a^\dag a}$, and
the shaded areas mark the basis states required to capture \SI{50}{\percent}, \SI{90}{\percent} or \SI{99}{\percent}
of the weights $\abs{\braket{n}{\psi(t)}}^2$ of the state.
\mname{} makes use of the fact that most states do not contribute at a given time and
that the set of states that \emph{do} contribute changes over time.}
\label{fig:moving_basis_set}
\end{figure}

As a simple illustrative example, consider the time evolution of a linearly displaced harmonic oscillator,
i.e., a system described by a Hamiltonian of the type
$H = \hbar\omega a^\dag a + \hbar \lambda \left(a + a^\dag\right)$,
from an initial vacuum state $\ket{\psi(0)} = \ket{0}$.
The time evolution (which corresponds to oscillating coherent states) is illustrated in Fig.~\ref{fig:moving_basis_set}:
We see that most of the Hilbert space is irrelevant to the time evolution at any given time and that a relatively small
set of basis states suffices to describe the majority of the state at any time,
but that the set of relevant basis states changes (or, in this case, oscillates) over time.
\begin{figure}
\centering
\includegraphics{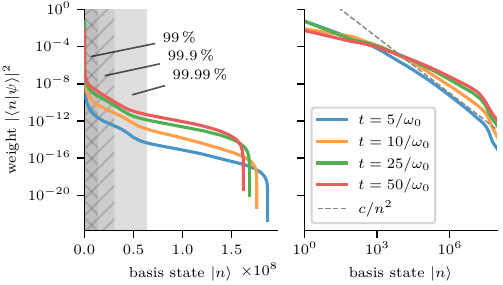}
\caption{The distribution of the expansion coefficients for a 1D Holstein model at several times after the start
of the evolution from a fully localized Franck--Condon excitation, in a lin-log plot (left) and a log-log plot (right).
We see that the vast majority of basis states contribute negligibly to the total wavefunction,
and the distribution largely seems to be between an exponential and a power law, spreading out over time.
The hatched areas apply to the fat-tailed $t = 50/\omega_0$ state and show that
less than half of the coefficients account for \SI{99.99}{\percent} of the total weight.
The dashed line shows a $1/n^2$ power law that is speculated to be attained here (see section~\ref{sec:benchmarking}).
Data is taken from the $g=4 \omega_0$ \mname{} calculation shown in Fig.~\ref{fig:benchmark}.}
\label{fig:coeff_distribution}
\end{figure}
Thus, at any time, we could imagine following the state in its time evolution with only 10 or 20 basis states
and capture essentially the entire dynamics by neglecting all the basis states outside of these basis states,
thereby saving enormous computational resources by not even constructing the irrelevant parts of the Hamiltonian, state vector, etc.
This concept is visualized in a more complicated and realistic scenario in Fig.~\ref{fig:coeff_distribution},
which shows the highly localized distribution of the coefficients $\abs{\braket{n}{\psi}}^2$
in a non-equilibrium Holstein model~\eqref{eq:holstein} that will be discussed in more detail in section~\ref{sec:benchmarking}.

If the underlying true $\mathcal{H}$ is infinite-dimensional (e.g., bosonic),
then a finite- but high-dimensional subspace $\mathcal{H}'$ must be chosen as a new underlying Hilbert space
within which \mname{} will then adaptively construct the low-dimensional effective Hilbert spaces.
The dimension of the space $\mathcal{H}'$ may be chosen to be quite large:
If a well-suited basis is used, then most basis states of $\mathcal{H}$ (and $\mathcal{H}'$) are not explored in finite time
and the effect of the dimension of $\mathcal{H}'$ is mostly restricted to a lookup table
that connects the truncated state vector with its physical index.
As a function of the total dimension $d_\text{tot} = \dim{\mathcal{H}'}$,
the memory required by this lookup table grows as $\mathcal{O}(\log_2{d_\text{tot}})$ (see App.~\ref{app:compression}).

As visualized in Fig.~\ref{fig:basic_idea},
such an adapted Hilbert subspace restricts the possible \enquote{movement} or propagation of the state vector at each timestep.
Though the time evolution can be performed to an essentially exact degree within the effective Hilbert space,
the timestep $\delta t$ must be chosen small enough---and the Hilbert space surrounding the current state large enough---%
so that the restriction to the effective subspace does not skew the results.
A greater problem that can arise is diffusion-like propagation or spreading of wavefunctions in a given basis:
Contrast the coherent behavior of a simple system such as the harmonic oscillator shown in Fig.~\ref{fig:moving_basis_set}
with the increasing spread of the wavefunction coefficients over time shown in the Holstein
model in Fig.~\ref{fig:coeff_distribution}.
In unfavorable situations, the entire state vector may at some point expand to a degree that exceeds the available memory,
leading to truncation artifacts which will be discussed in more detail in section~\ref{sec:performance}.

\section{Details of \mname{}}\label{sec:details}
We shall now present the building blocks that form the foundation of the \mname{} method,
beginning with the basics of its underlying vector representation.

\subsection{When is a vector-based representation useful?}\label{sec:info_comparison}
Before exploring in detail how the method treats the time evolution of states,
we will demonstrate under which conditions the vector-based state representation underpinning \mname{} or may not be advantageous.
Beyond the cost of storing the expansion coefficients themselves,
there is also a memory cost associated with storing information on the basis states themselves,
explained in detail in App.~\ref{app:compression}.
The following discussion will focus on the expansion coefficients first and then factor in this additional cost afterwards.

As MPS methods are some of the most widespread numerically exact techniques in condensed-matter physics,
in part because of their often highly memory-efficient representation of states,
we will compare the memory required to store four example states exactly using either \mname{} or a standard MPS approach.
Given that we refer to them frequently in the following, a basic overview of MPS with a particular focus on their
relevance to the presence is presented in App.~\ref{app:mps}---%
we also point out that the multilayer variant of MCTDH, i.e., ML-MCTDH~\cite{Wang2003JCP,Wang2009Wiley,Wang2015},
has recently begun to be understood as a tree-tensor-network~\cite{Larsson2024MolPhys,Dorfner2024JCTC},
of which matrix product states are a particular one-dimensional variant,
meaning that some of the comparisons made here also carry over to ML-MCTDH.

All of the following example states shall be states in a Hilbert space $\mathcal{H} = \mathcal{H}_\text{qudit}^{\otimes L}$
describing a length-$L$ chain of $d$-dimensional qudits.

\subsubsection*{Example 1: Highly entangled states}
As a first example case,
consider the generalized GHZ state
\[\ket{\psi_\text{GHZ}} =  \frac{1}{\sqrt{d}} \sum_{k=0}^{d-1} \ket{kk \cdots kk}.\]
The number of non-zero expansion coefficients needed to store this state in \mname{} is simply the local dimension $d$,
independent of the number of sites $L$.
This is the case despite the well-known fact that such states are highly entangled:
Along any bipartition of the lattice, this state has entanglement $\log_2{d}$,
and therefore its MPS representation requires a bond dimension~\cite{Eisert2010} of $\chi_\text{max} = d$.
Consequently, the central tensor alone requires storing a total of $d \times d \times d$ matrix elements.
One thus finds that, potentially surprisingly, the relatively naïve vector representation is much more efficient in this situation.

\subsubsection*{Example 2: Delocalization \& coherence}
Of course, this is not always the case.
An almost trivial example of how MPS compression can beat wavefunction truncation is the following state,
\[\ket{\psi_\text{delocal}} = \frac{1}{d^{L/2}} \sum_{k,\ldots,k_L} \ket{k_1 \ldots k_L},\]
which could be considered \enquote{completely delocalized} in the given basis.
It is exactly representable as an MPS with bond dimension $\chi_\text{max} = 1$ using the $1 \times 1$-matrix
$A_{k_i} = \begin{pmatrix} 1 /\sqrt{d} \end{pmatrix}$
for every lattice site $i$ and basis state $k$.
This is, naturally, due to the fact that this state is completely unentangled; it can be rewritten as a product state
\[\ket{\psi_\text{delocal}} = \left( \sum_{k_1} \frac{1}{\sqrt{d}} \ket{k_1} \right) \otimes 
\cdots \otimes  \left( \sum_{k_L} \frac{1}{\sqrt{d}} \ket{k_L} \right),\]
which is virtually the best-case scenario for MPS.
A sparse-vector method, on the contrary, would maximally struggle with such a state,
as the entire basis set is required to represent it exactly (and, furthermore,
due to the complete degeneracy of the expansion coefficients, every choice for a truncation would be equally bad).
However, this deficiency is completely basis-dependent, as we could unitarily rotate the single-site basis such that
\[\ket{0'} \coloneqq \sum_{\sigma = 0}^{d-1} \frac{1}{\sqrt{d}} \ket{k}.\]
Then the same state, in the new basis, reads
$\ket{\psi_\text{delocal}} = \ket{0'}^{\otimes L}$,
which is exactly representable by a single expansion coefficient---%
but such basis transformations often have undesirable side effects for the time evolution,
as they may undo the sparseness of the Hamiltonian that will play an important role in the following sections.

\subsubsection*{Example 3: Large local dimensions}
Going beyond the salient features of entanglement and superposition highlighted in the first two examples,
the more subtle third example analyzes the deleterious effects of (unnecessarily) high-dimensional Hilbert spaces
using the very simple state
$\ket{\psi_0} = \ket{00\cdots 00}$.
This requires only a single expansion coefficient to be stored,
but now the aforementioned cost of storing the lookup table becomes important,
as it penalizes large Hilbert spaces irrespective of the current state vector.
As explained in App.~\ref{app:compression},
each row in the lookup table of a $d^L$-dimensional space requires $L \log_2{d}$ bits of memory
(plus, potentially, some mostly negligible overhead of less than the size of a single integer in case of incommensurate values).

As an MPS, this simple state also requires no more than $\chi_\text{max} = 1$---%
but a similar issue as described above also arises here,
since it is still necessary to store $L$ third-order tensors of dimension $d \times 1 \times 1$ each,
or $d L$ numbers in total. 
Regardless of the method, this behavior is especially disadvantageous if $d$ or $L$ is larger than necessary, i.e.,
if $d$ or $L$ can be reduced without affecting the observable dynamics.

\subsubsection*{Example 4: Long-range entanglement}
Finally, consider the generalized NOON state:
\[\ket{\psi_\text{NOON}} =  \frac{1}{\sqrt{2}} \left(\ket{00\cdots 01} + \ket{10\cdots 00}\right).\]
This can also be seen as a Bell state separated over a large \enquote{distance}.
To the vector-based representation, this is simply a superposition of two states,
equivalent in memory requirements to a local superposition such as
$\frac{1}{\sqrt{2}} \left(\ket{0} + \ket{1}\right) \otimes \ket{0}^{\otimes (L-1)}$.
However, the $\log_2{2}$ entanglement necessitates an MPS bond dimension of $\chi_\text{max} = 2$ on every site:
Thus, in order to pipe this weak entanglement through the entire MPS,
it is necessary to spend memory at every site connecting the two.

This last example may seem contrived and unphysical, as one may argue that long-range correlations are rather rare
and would realistically be subject to considerable decoherence in natural systems.
However, situations such as this may indeed arise in practice if the system is not
geometrically one-dimensional and it was necessary to force a linear ordering upon a system:
For example, two neighboring sites in 3D are not necessarily neighboring when ordered in 1D,
leading to seemingly long-ranged entanglement.

\subsubsection{Implications and limitations}
\begin{table}
\caption{The memory required to represent each state exactly
using either matrix-product states or \mname{},
including the space required to store the corresponding lookup table in \mname{},
assuming $L \log_2{d}$ is commensurate with the lookup table's wordsize and using 128-bit complex number for the coefficients.
Results are stated in units of 128 bits.}
\begin{center}
\begin{tabular}{l@{\quad}l@{\quad}l}
\toprule 
state                  	        & matrix-product state  & sparse vector                                 \\\midrule
$\ket{\psi_\text{GHZ}}$       	& $2d^2 + (L-2)d^3$     & $d \left(1 + \frac{L\log_2{d}}{128}\right)$   \\
$\ket{\psi_\text{delocal}}$     & $dL$                  & $d^L \left(1 + \frac{L\log_2{d}}{128}\right)$ \\
$\ket{\psi_0} $                 & $dL$                  & $1 + \frac{L\log_2{d}}{128}$                  \\
$\ket{\psi_\text{NOON}}$        & $4d(L-1)$             & $2 + \frac{L\log_2{d}}{64}$                   \\
\bottomrule
\end{tabular}
\end{center}
\label{tab:memory_comparison}
\end{table}
A quantitative comparison of the above examples, including the cost of the lookup table for \mname{},
is given in Table~\ref{tab:memory_comparison}.
As expected, MPS representations work very well for states with low or locally confined entanglement,
whereas the representation used by \mname{} naturally excels in situations in which the given basis
leads to a sparse representation of the state vector:
Coherence (in the given basis), not entanglement, then becomes the deciding factor for efficient representations.

Two important consequences of our analysis are:
First, since entanglement is invariant under local unitary transformations~\cite{Mintert2009},
this limitation of MPS cannot be remedied (but also not exacerbated) by a local change of basis,
whereas the vector truncation method is affected by a the choice of the local basis.
However, \emph{non-local} basis transformations can affect both methods:
If such a transformation decreases long-range entanglement, MPS methods can be expected to perform better,
whereas a (non-local or local) basis transformation that concentrates the wavefunction on a smaller number of basis states will benefit \mname{}.
Second, both MPS representations and vector-based representations suffer when used with unnecessarily large local dimensions,
as it then becomes necessary to either drag along many unnecessary physical indices (in the case of MPS)
or to expend memory on a very large lookup table (in the case of sparse vectors),
though the latter scales only logarithmically with the total Hilbert-space dimension.
It should be noted that there have been efforts to counter the issues encountered by MPS in these examples,
for example by employing sparse tensors~\cite{Hubig2017PRB},
by exploring the impact of reordering the system indices~\cite{Lacroix2025,Lacroix2025arXiv}
or employing methods such as local basis optimization~\cite{Brockt, LBO1, LBO2} (also discussed in App.~\ref{app:mps}),
though such approaches may require investing additional resources to solve the problem.

In summary, the suitability of the given basis can make or break
the ability of a truncated-vector method to represent a given state and subsequently the feasibility of \mname{}.
However, finding the optimal basis for the present method requires finding a delicate balance between
the complexity of the dynamics and the sparseness of the Hamiltonian matrix.
For example, the Lang--Firsov or polaron transformation~\cite{LangFirsov,Xu2016Frontiers}
may be beneficial to solving the Holstein Hamiltonian~\eqref{eq:holstein}
in that it locally transforms the problem such that off-diagonal elements of the Hamiltonian become smaller,
but it also makes the Hamiltonian much less sparse due to quasi-non-local couplings in the new basis.

Though the examples given above may be illustrative, they have so far only considered exact representations of a single state:
For one, systematic investigation and comparison of the loss of information when a state is \emph{compressed}
in either of the two schemes would be interesting and relevant for future developments.
Furthermore, this analysis has not covered how the suitability of each representation changes as the state is evolved,
which, for MPS methods, primarily depends on buildup of correlations and entanglement between sites,
while \mname{} depends primarily on the connectivity of the Hamiltonian,
which may cause favorable states to evolve into unfavorable ones---%
effects which will be explored in more detail in the following sections.
Finally, a highly relevant question is which of these example states is actually most likely
to be representative of those that are encountered \enquote{in the wild}.
Though the preceding analysis has commented on when a given scenario may or may not be likely,
the true state will depend on the system at hand and its behavior cannot necessarily be accurately predicted in advance.

\subsection{Sparse Hamiltonians}\label{sec:sparseHam}
Most Hamiltonians encountered in condensed-matter physics, when written in a \enquote{reasonable} basis,
are sparse due to the local nature of the interactions:
That is to say, the number of non-zero entries in the Hamiltonian matrix is far smaller than the number of vanishing entries.

For example, when representing the single-exciton 1D Holstein Hamiltonian~\eqref{eq:holstein}
as a matrix even with a very small fixed truncated phononic dimension $d_\text{pho} = 5$ and chain length $L = 9$,
one finds that less than $\frac{1}{\num{4000000}}$ of the matrix elements are non-zero,
and increasing the system size in either dimension further reduces density of the Hamiltonian.
More details and another example based on spin systems are given in App.~\ref{app:sparse}.

\mname{} makes use of this property in two separate ways.
For one, the sparse nature of typical Hamiltonians can be exploited to represent them in dedicated sparse matrix formats.
The use of sparse matrices requires saving some additional information about the position of a certain value in the matrix,
but for all but the smallest or relatively dense matrices, the memory savings still outweigh the overhead.
The sparseness (and, more generally, block-sparseness) of many Hamiltonians has been widely recognized
in the many-body community, including in many implementations of MPS methods~\cite{Hubig2017PRB,Keller2015JCP,Ren2020JCP,Bachmayr2022Calcolo,vanDamme2024SciPost,Rams2025SciPost},
as it can lead to significant improvements in efficiency if effectively utilized.
With regard to \mname{},
it should be pointed out that a sparse Hamiltonian matrix $H$ often still leads to a dense propagator matrix
$U(\delta t) = \exp(-i H \delta t/\hbar)$, even for short $\delta t$.
This issue can be effectively circumvented by never explicitly constructing the propagator matrix $U(\delta t)$,
but instead only evaluating its action on the current state vector $\ket{\psi}$, as described in App.~\ref{app:propagator}.

However, beyond the relatively simple fact that many Hamiltonians are well-suited to being represented as sparse matrices,
the sparse nature of typical Hamiltonians plays another critical role in \mname{}.
This is because the method depends crucially on closely related property of a given Hamiltonian:
its \emph{connectivity}, that is, how many other basis states $\{\ket{k}, \ket{l}, \ldots \}$
a given basis state $\ket{m}$ is mapped onto by the Hamiltonian---%
in graph theory, this is known as the \emph{degree} of a vertex~\cite{Wilson1996}.
This, in turn, is closely related to the concept of neighbor states that will be described in the following section,
where we will also detail how this property ultimately decides how effective \mname{} is at handling a given problem.

\subsection{Co-evolving subspaces}\label{sec:adaptation}
We have alluded to the concept of \enquote{neighboring} basis states several times
and mentioned how they are used to span the adapted effective Hilbert space.
We will now describe how these neighboring basis states are determined.

We first define two example Hamiltonians that we will frequently refer to in the following:
The 1D single-particle tight-binding (TB) Hamiltonian $H_\text{TB}$ with open boundary conditions, 
\begin{equation}
    H_\text{TB} = \sum_{j=1}^L \epsilon_j \dyad{j}
        + \hbar \left[\sum_{j=1}^{L-1} J_j \dyad{j}{j+1} + \text{h.c.}\right],
    \label{eq:tb}
\end{equation}
and the (Frenkel-)Holstein Hamiltonian~\cite{HolsteinPart1,Hestand2018ChemRev},
which can be viewed as an extension of the TB Hamiltonian:
\begin{equation}
    H_\text{H} = H_\text{TB}
        + \hbar \sum_{j=1}^L \left[ \omega_j {a_j}^\dag a_j + g_j \dyad{j} \left({a_j}^\dag + a_j\right)\right],
    \label{eq:holstein}
\end{equation}
where $a_j$ is the annihilation operator of the $j$-th quantum harmonic oscillator (QHO)
representing a vibrational (phononic) mode at site $j$, and $\ket{j}$ is the excitonic wavefunction when the exciton
is fully localized to the $j$-th site.
$\omega_j$, $\epsilon_j$, $g_j$ and $J_j$ are the $j$-th phonon frequency, local exciton energy, exciton--phonon coupling
and excitonic hopping strength (nearest-neighbor dipole-dipole coupling), respectively.
In the following, the parameters are often identical for each site, in which case we will drop the index $j$
(except for the QHO frequency $\omega$, which is instead called $\omega_0$ to avoid confusion
with the Fourier-transform frequency in section~\ref{sec:aggregates}).
Note that due to the restriction to the single-particle manifold,
the model can be interpreted as either a single-electron or a single-exciton model without further loss of generality.

\subsubsection{Motivation}
As a simple example, consider a single-particle tight-binding chain~\eqref{eq:tb} of infinite length:
The wavefunction of an initially localized particle in such a system will spread ballistically over time.
Assume that at some point in time, the state is delocalized over the five basis states
\[\{\ket{n-2}, \ket{n-1}, \ket{n}, \ket{n+1}, \ket{n+2}\}.\]
As a very crude approximation, we could restrict the effective Hilbert space for the next timestep
to the span of only these five states, but doing so would introduce obvious restriction artifacts.
Clearly, we should include the next few sites in the chain so that the system can freely evolve into the neighboring states
while still neglecting the very far-away basis states which will not be occupied within the next timestep.
However, though this concept of neighboring states is evident in a well-ordered chain,
how should we determine which basis states are \enquote{neighbors} in a basis without an obvious ordering?

\subsubsection{Defining neighboring basis states} \label{sec:neighbors}
Since we are interested in time evolution, which is ultimately governed by the Hamiltonian,
a reasonable answer to this question is to consider the structure that the Hamiltonian itself induces on the basis states.
Let $\mathfrak{B}$ be the orthonormal basis in which we are representing our vectors and matrices,
with $\spn{\mathfrak{B}} = \mathcal{H}$ the total Hilbert space.
As a linear operator, the Hamiltonian maps each basis state $\ket{n} \in \mathfrak{B}$ onto a superposition
\begin{align*}
H\ket{n} &= \sum_{\ket{k} \in \mathfrak{B}} h_{jn} \ket{j} \\
    &= \sum_{\ket{m} \in \mathcal{N}_{\ket{n}}} h_{mn} \ket{m}
    + \sum_{\ket{j} \in \mathfrak{B} \setminus \mathcal{N}_{\ket{n}}} 0 \ket{j},
\end{align*}
where the set of basis states $\mathcal{N}_{\ket{n}}$ corresponds to the subset of all basis states in $\mathcal{H}$
that $\ket{n}$ is mapped onto by $H$. Symbolically:
\[\mathcal{N}_{\ket{n}} \coloneqq \left\{\ket{m} \in \mathfrak{B} \mid \mel{m}{H}{n} \neq 0 \right\}.\]
This definition readily extends to a state $\ket{\psi}$ which is not a basis state,
or even an arbitrary set $\mathcal{M}$ of basis states (which may also be identified with the subspace it spans):
\begin{align}\begin{split}
\mathcal{N}_{\ket{\psi}} &\coloneqq \left\{\ket{n} \in \mathfrak{B} \mid \mel{n}{H}{\psi} \neq 0 \right\},\\
\mathcal{N}_{\mathcal{M}} &\coloneqq \left\{\ket{n} \in \mathfrak{B} \mid \left(\exists \ket{m} \in \mathcal{M}\right)\left[\mel{n}{H}{m} \neq 0\right] \right\}.\label{eq:def_neighbor}
\end{split}\end{align}
We can now use this definition to call $\mathcal{N}_{\mathcal{M}}$ the \emph{neighboring states of $\mathcal{M}$}.
Note that $\ket{n}$ may be contained in $\mathcal{N}_{\ket{n}}$, for example when $H$ has a non-zero diagonal element $h_{nn}$---%
if this is the case, call $\ket{n}$ a \emph{trivial neighbor} of $\mathcal{M}$.
We can also easily generalize~\eqref{eq:def_neighbor} to \emph{higher-order} neighbors:
\begin{equation}
\mathcal{N}_{\mathcal{M}}^{(k)} \coloneqq \left\{\ket{n} \in \mathfrak{B} \mid \left(\exists \ket{m} \in \mathcal{M}\right)\left[\mel{n}{H^k}{m} \neq 0\right] \right\}.\label{eq:def_neighbor_k}
\end{equation}
The set $\mathcal{N}_{\mathcal{M}}^{(k)}$ then denotes the set of basis states into which
the set of states $\mathcal{M}$ can evolve under $k$-fold application of the Hamiltonian.
Furthermore, note that $\mathcal{N}_{\mathcal{M}}^{(0)} = \mathcal{M}$.

Let us illustrate this concept for first-order neighbors, $k = 1$, using the TB model~\eqref{eq:tb}.
Let $\mathcal{M}$ be the set consisting of the two basis states
\[\mathcal{M} = \{\ket{n-1}, \ket{n+1}\}.\]
The TB Hamiltonian with a constant hopping parameter $J_k = J$ and zero on-site energy $\epsilon_k = 0$ maps the first state $\ket{n-1}$ onto
\[\ket{n-1} \longmapsto J\left( \ket{n-2} + \ket{n} \right),\]
and the Hamiltonian maps the second state onto 
\[\ket{n+1} \longmapsto J\left( \ket{n} + \ket{n+2} \right).\]
Therefore, in this case the neighbor set is
\[\mathcal{N}_{\mathcal{M}} = \left\{\ket{n-2}, \ket{n}, \ket{n+2}\right\},\]
which is precisely what one would intuitively expect when speaking of the neighboring lattice sites of $\{\ket{n-1}, \ket{n+1}\}$. If the on-site energy is non-zero, $\epsilon_{n \pm 1} \neq 0$, then $\ket{n-1}$ is also mapped to itself and
and similarly for $\ket{n+1}$, such that the neighbor set becomes
\[\mathcal{N}_{\mathcal{M}} = \left\{\ket{n-2}, \ket{n-1}, \ket{n}, \ket{n+1}, \ket{n+2}\right\},\]
where we can see that $\mathcal{M} \cap \mathcal{N}_{\mathcal{M}} = \mathcal{M}$,
i.e., $\mathcal{N}_{\mathcal{M}}$ now contains trivial neighbors.

The effective Hilbert space that is used to evolve $\ket{\psi}$ from time $t$ to $t + \delta t$ is constructed as
\begin{equation}
\mathcal{H}_\text{eff}^{(m)}(t)
    = \spn\fleft\{\bigcup_{k=0}^{m} \mathcal{N}_{\ket{\psi(t)}}^{(k)}\fright\}.
\label{eq:H_eff}
\end{equation}
Clearly, $\mathcal{H}_\text{eff}^{(m)}(t)$ in the limit $m \to \infty$ becomes the entire Hilbert space---%
or at least the entire Hilbert subspace that is reachable from the given initial state due to symmetries---%
but increasing $m$ indefinitely is not practically possible.
In practice, a large maximum neighbor order $m_\text{init}$ is used to generate
the initial $\mathcal{H}_\text{eff}^{(m_\text{init})}(0)$ surrounding the initial state $\ket{\psi(0)}$,
but a smaller $m$ is chosen when dynamically adapting at $t > 0$:
Concrete examples will be shown in sections~\ref{sec:benchmarking} and \ref{sec:aggregates}.
Note that we will drop the superscript $(m)$ on $\mathcal{H}_\text{eff}^{(m)}(t)$ of~\eqref{eq:H_eff} when its value is not important.

An important and useful feature of this approach should be noted:
If the Hamiltonian $H$ under consideration induces a symmetry $S$ of the system ($\comm{H}{S} = 0$)
and the state $\psi$ that is being evolved lies fully within one symmetry sector
($S\ket{\psi} = \lambda \ket{\psi}$ where $\lambda$ is some number),
then the neighbor-states approach described here will automatically respect that symmetry
and never include basis states belonging to a different symmetry sector.
For example, the effective Hilbert spaces constructed for a state with a definite particle number that is evolving
under a particle-number-conserving Hamiltonian will automatically be restricted to the appropriate particle-number subspace.
This behavior can potentially greatly increase numerical efficiency without any further user input.
Such behavior may also be realized, for example, in MPS methods,
but requires additional effort to be invested in symmetrization~\cite{Singh2010PRA,Paeckel2017SciPost,Rams2025SciPost}.

Furthermore, we point out that the series-based time-evolution operator $U(\delta t)$ inside $\mathcal{H}^{(m)}_\text{eff}$
is typically expanded to a much higher degree than the maximal neighbor order $m$ (see App.~\ref{app:propagator}):
This allows for virtually unitary and exact time evolution \emph{within the restricted subspace}.
Since the evaluation of the time evolution operator as described in App.~\ref{app:propagator} is very efficient
in both computational time and memory, it is expedient to continue the series expansion of $U(\delta t)$ until
the single-timestep evolution is fully converged.
It is important to note that such a high-order expansion within $\mathcal{H}_\text{eff}^{(m)}$
(where $m$ is much smaller than the order of the expansion)
is \emph{not} equivalent to an $m$-th order expansion of the Taylor series, which would quickly become non-unitary
and provide unreliable results beyond a short initial time frame (demonstrated in App.~\ref{app:Taylor}).
We also point out that, though an $m$-th-order neighboring subspace guarantees that all neighboring states
up to order $m$ are \emph{included}, it does \emph{not necessarily exclude} all neighbors of order $m+1$ and higher,
as the set of basis states comprising a typical state will already contain many of its own neighboring basis states
(i.e., the support of a typical state already contains many trivial neighbor states).

\subsubsection{The Hamiltonian as a graph and its connectivity}
On a theoretical level, this procedure of \enquote{neighbor-finding} induces
a (weighted) graph on the basis states of the Hilbert space.
This Hamiltonian-induced graph structure is precisely what determines
how closely or distantly different basis states are related for our purposes.
This line of thought has previously been used in the study of continuous-time quantum walks~\cite{Farhi1998PRA}
and thermalization and ergodicity in quantum systems~\cite{Roy2020PRR,Desaules2022PRB,Menzler2025PRB}.

In the same vein, one finds that a central quantity underpinning the efficacy of \mname{} is the \emph{connectivity}, which we define as
\begin{equation}
\kappa_{\mathcal{M}}^{(k)} \coloneqq \frac{ \abs{\mathcal{N}_{\mathcal{M}}^{(k)}} }{ \abs{\mathcal{M}} }
\qq{or}
\kappa_{\ket{\psi}}^{(k)} \coloneqq \frac{ \abs{\mathcal{N}_{\ket{\psi}}^{(k)}} }{ \abs{\mathcal{N}_{\ket{\psi}}^{(0)}} }, \label{eq:connectivity}
\end{equation}
where $\abs{\mathcal{N}_{\ket{\psi}}^{(0)}} = \left\{\ket{n} \in \mathfrak{B} \mid \braket{n}{\psi} \neq 0 \right\}$
and $\abs{\cdot}$ is the magnitude of a set.
For an isolated basis state, $\kappa_{\ket{m}}^{(1)}$ is exactly equal to the graph-theoretical \emph{degree}
of the vertex $\ket{m}$ in the graph induced by the Hamiltonian~\cite{Wilson1996}.
This quantity is crucial to \mname{} because it simultaneously determines how much $\mathcal{H}_\text{eff}^{(m)}$ of~\eqref{eq:H_eff}
will grow with increasing $m$ and is also intimately related to how much the wavefunctions will expand within a subspace.
A low connectivity is therefore central to efficient and well-converged results with \mname{},
and, as alluded to in section~\ref{sec:sparseHam}, is also closely tied to the sparseness of the Hamiltonian.
However, as we will see in sec.~\ref{sec:aggregates} and App.~\ref{app:conv_spectra},
even systems with identical connectivity do not necessarily behave equally well under \mname{},
revealing that the connectivity is not the only relevant factor.

\subsubsection{Relation to Krylov subspaces}\label{sec:Krylov}
The concept of
$\mathcal{H}_\text{eff}^{(m)} = \spn\fleft\{\bigcup_{k=0}^{m} \mathcal{N}_{\mathcal{M}}^{(k)}\fright\}$
of~\eqref{eq:H_eff} may at first seem similar to the Krylov subspace $K^{m+1}$,
defined as~\cite{Krylov,Saad2003}
\[ K^{m+1}(H, v) = \spn\fleft\{ v, Hv, H^2v, \ldots, H^m v\fright\}.\]
The two are, however, quite different:
This can be easily seen by considering that the Krylov subspace $K^{m+1}$ is (at most) an $(m+1)$-dimensional vector space,
whereas the dimension of $\mathcal{H}_\text{eff}^{(m)}$ is typically far larger than $m$
and depends on both the basis in which the system is expressed as well as the nature of both the Hamiltonian $H$ and the vector $v$.
For example, consider the state
\begin{equation}
\ket{\psi} = \frac{1}{\sqrt{2}} \left(\ket{n-1} + \ket{n+1}\right) \label{eq:psi_krylov},
\end{equation}
on an infinite TB chain~\eqref{eq:tb} with zero on-site energy $\epsilon = 0$.
We find that
\[
\abs{\mathcal{N}_{\ket{\psi}}^{(0)}} = 2, \qquad
\abs{\mathcal{N}_{\ket{\psi}}^{(1)}} = 3, \qquad
\abs{\mathcal{N}_{\ket{\psi}}^{(2)}} = 4,
\]
leading to an effective Hilbert space $\mathcal{H}_\text{eff}^{(2)}$ spanned by
\[
\mathcal{N}_{\ket{\psi}}^{(0)} \cup \mathcal{N}_{\ket{\psi}}^{(1)} \cup \mathcal{N}_{\ket{\psi}}^{(2)}
    = \big\{\!\ket{n-3}, \ket{n-2}, \ldots, \ket{n+3}\!\big\}.
\]
The spanned vector space is seven-dimensional, where the subadditive nature is due to
$\mathcal{N}_{\ket{\psi}}^{(0)} \subsetneq \mathcal{N}_{\ket{\psi}}^{(2)}$.
On the other hand, the corresponding Krylov subspace $K^3$ is always three-dimensional:
\begin{multline*}
K^3(H_\text{TB}, \ket{\psi}) = \spn\!\Big\{\!\left(\ket{n-1} + \ket{n+1}\right),\\
\left(\ket{n-2} + 2\ket{n} + \ket{n+2}\right), \\
\left(\ket{n-3} + 3\ket{n-1} + 3\ket{n+1} + \ket{n+3}\right)\!\Big\},
\end{multline*}
clearly a quite different result from the corresponding neighbor-states construction $\mathcal{H}_\text{eff}^{(2)}$.

Furthermore, though $K^{m+1}$ is always contained as a subspace in $\mathcal{H}_\text{eff}^{(m)}$ by construction,
the converse is not true: In the example~\eqref{eq:psi_krylov} given above, the state $\ket{n}$ is not contained in $K^3$
but \emph{is} contained in $\mathcal{H}_\text{eff}^{(2)}$.
More generally, even though higher-order Krylov spaces such as $K^{m+2}$ will typically be spanned by basis states
that do not fully lie in $\mathcal{H}_\text{eff}^{(m)}$, the additional information conferred by increasing
the dimension of the Krylov space is oftentimes partially contained inside a lower-order $\mathcal{H}_\text{eff}^{(m)}$.
For the example state~\eqref{eq:psi_krylov}, the orthogonal vector $v_4$ that would be added by going from
the three-dimensional Krylov subspace $K^3$ to the four-dimensional $K^4$ is proportional to (after orthogonalization)
\[v_4 \propto \ket{n - 4} + \frac{2}{3}\Bigl(\ket{n - 2} - \ket{n} + \ket{n+2}\Bigr) + \ket{n + 4}.\]
Despite the basis states $\ket{n \pm 4}$ not being contained in $\mathcal{H}_\text{eff}^{(2)}$,
it turns out that $\mathcal{H}_\text{eff}^{(2)}$ is still far from orthogonal to this additional dimension of $K^4$,
as the unit vector $\frac{1}{\sqrt{3}} \left(\ket{n - 2} - \ket{n} + \ket{n + 2}\right) \in \mathcal{H}_\text{eff}^{(2)}$
has an overlap of $\sqrt{2 / 5} \approx 0.63$ with $v_4$.
Likewise, though the odd-numbered higher-order orthonormalized Krylov vectors $v_5, v_7, \ldots$
\emph{are} orthogonal to all states in $\mathcal{H}_\text{eff}^{(2)}$ in this example,
the maximal overlap between the unit sphere in $\mathcal{H}_\text{eff}^{(2)}$ and the even-numbered Krylov vectors
$v_6$ and $v_8$ is $\sqrt{6/35} \approx 0.41$ and $\sqrt{2/21} \approx 0.31$, respectively.
This indicates that an effective Hilbert space $\mathcal{H}_\text{eff}^{(2)}$
as constructed in~\eqref{eq:H_eff} with $m = 2$ is partially capable of capturing dynamics contained in
higher-order Krylov subspaces $K^{m' + 1}$ with $m' > 2$,
as is further demonstrated numerically for the 1D Holstein model in App.~\ref{app:Lanczos}.
Note that these effects are attributable to the occurrence of trivial neighbors within a given basis set,
and the $(n+1)$-th Krylov vector $v_{n+1}$ (arising from the $n$-fold application of $H$)
is contained fully within $\mathcal{H}_\text{eff}^{(n-1)}$
if all basis states spanning $\mathcal{H}_\text{eff}^{(n-1)}$ are trivial neighbors,
resulting in $\mathcal{H}_\text{eff}^{(n)} = \mathcal{H}_\text{eff}^{(n-1)}$.
This exact closure, however, is much less likely to occur than the \enquote{partial} closure
demonstrated with the TB example above. 

Having concluded that the effective Hilbert spaces constructed by the present method are, though related,
not equivalent to Krylov subspaces, one may nevertheless wonder whether employing a Krylov-based Lanczos procedure
for the single-timestep evolution could be beneficial.
One downside of a Lanczos procedure in this situation is memory:
Depending on the order of the Krylov expansion, 
the additional memory requirements of storing the change-of-basis matrix that maps
between the high-dimensional $\mathcal{H}_\text{eff}$ and the low-dimensional Krylov subspace may be much more costly
than the series-based propagation used by the current implementation of \mname{}, which in turn then also
outweighs potential time savings gained by performing the time-evolution step within the smaller Krylov subspace.
This is described and demonstrated in more detail in App.~\ref{app:Lanczos}.

\subsubsection{Truncation \& post-adaptation detruncation}\label{sec:detruncation}
\mname{} uses~\eqref{eq:H_eff} to construct the effective Hilbert space $\mathcal{H}_\text{eff}(t)$
based on a preset maximum neighbor order $m$.
However, if left to its own devices, $\mathcal{H}_\text{eff}(t)$
will typically grow with every timestep as $\ket{\psi(t)}$ continues to delocalize,
soon exceeding the available computational memory.
Therefore, the set of basis states must be periodically truncated:
Empirically, the best truncation procedure so far seems to be simply weighting the basis states
according to the magnitude of their current expansion coefficient, though other approaches are also conceivable
(see App.~\ref{app:truncation} for further discussion).

In its current implementation, \mname{} truncates the set of basis states at each timestep
prior to \enquote{re-growing} the Hilbert space via~\eqref{eq:H_eff}.
In principle, this would discard the expansion coefficients associated with any truncated basis state,
but there is a subtle yet important trick that can be used in the interplay between truncation and adaptation:
After applying~\eqref{eq:H_eff} and determining the new effective Hilbert space,
one can \enquote{rescue} any possible values that had originally been marked for truncation
but whose basis states were reinstated by the adaptation.
More specifically: Say we want to truncate the state vector $\ket{\psi(t)}$
to the \qnom{} states that make up some set of basis states $\mathcal{M}$.
This set of states $\mathcal{M}$ is then fed into the adaptation algorithm which will produce
the adapted $\mathcal{H}_\text{eff}(t)$ as given by~\eqref{eq:H_eff}.
The dimension of $\mathcal{H}_\text{eff}(t)$ is some number $q_\text{true} \geq q_\text{nom}$.
Now, instead of actually keeping only the \qnom{} most important expansion coefficients of the vector,
i.e., instead of projecting $\ket{\psi(t)}$ to $\spn{\mathcal{M}}$,
we instead project $\ket{\psi(t)}$ to the larger Hilbert space $\mathcal{H}_\text{eff}(t) \supset \spn{\mathcal{M}}$,
thereby (typically) truncating to a lesser extent than expected and keeping up to \qtrue{} expansion coefficients.
In extreme cases, it is even possible that this \enquote{post-adaptation detruncation} results in no truncation at all,
if $\mathcal{H}_\text{eff}(t-\delta t)$ is entirely contained within $\mathcal{H}_\text{eff}(t)$---%
this corresponds to the exact closure mentioned in sec.~\ref{sec:Krylov}.

Furthermore, we can quantify the ratio between \qtrue{} and \qnom{} using the connectivity defined in~\eqref{eq:connectivity}:
\begin{equation}
    q_\text{true} = \kappa_{\mathcal{M}}^{(m)} q_\text{nom}. \label{eq:qtrue}
\end{equation}
Note, however, that even if each basis state has a constant connectivity $\kappa$ under the given Hamiltonian,
\qtrue{} will almost always be smaller than $\kappa^m q_\text{nom}$
due to the fact that most sets of states contain many trivial neighbors.
In other words, only the connectivity of the basis states that are currently at the \enquote{edge} of $\mathcal{H}_\text{eff}(t)$
is important to the growth of $\mathcal{H}_\text{eff}(t)$.
In the 1D Holstein system, the relationship between \qtrue{} and \qnom{} appears to be approximately given by
$q_\text{true} \sim 2^{m-1} 2.5 q_\text{nom}$:
For example, in the calculations shown in Fig.~\ref{fig:coeff_distribution},
the parameter \qnom{} was set to \num{32e6} and $m = 2$, but we see that \qtrue{} is approximately \num{170e6}.

\subsubsection{Generating the effective Hamiltonian from the effective Hilbert space}
As the dimension of the underlying Hilbert space is typically extremely large (easily exceeding $10^{50}$),
it is not possible to explicitly store the Hamiltonian in the underlying basis, even as a sparse matrix.
Therefore, a matrix-free method must be used to compute the action of the Hamiltonian used in the generation
of $\mathcal{H}_\text{eff}(t)$ of~\eqref{eq:H_eff}.
However, once $\mathcal{H}_\text{eff}(t)$ has been established,
the Hamiltonian inside of this lower-dimensional subspace is constructed as an explicit sparse matrix,
allowing for the use of efficient SpMV algorithms:
During the last application of the matrix-free Hamiltonian method used in constructing~$\mathcal{H}_\text{eff}(t)$,
we construct the Hamiltonian \emph{inside}~$\mathcal{H}_\text{eff}(t)$ by first retaining the association between
each basis state $\ket{n}$ and its image under the Hamiltonian 
and then also generating the matrix element associated with each such pair.
Combined with their Hermitian conjugates, the set of all such matrix elements
must add up to \emph{every} Hamiltonian matrix element within the effective subspace.
However, we need to take care to avoid creating duplicate entries of matrix elements, which can occur,
e.g., either due to the intersection of two neighbor sets being non-empty
($\mathcal{N}_{\mathcal{M}}^{(k)} \cap \mathcal{N}_{\mathcal{M}}^{(l)} \neq \emptyset$),
or due to the existence of non-trivial neighbors within $\mathcal{M}$.

\subsection{Formal operator representation and notes on parallelization}\label{sec:parallelization}
We can now give a more precise description of the procedure that was outlined in section~\ref{sec:the_essence}:
To evolve a state $\ket{\psi(t)}$ with a sparse representation in the basis $\mathfrak{B}$ forward by one timestep:
\begin{enumerate}
\item First construct the effective Hilbert space $\mathcal{H}_\text{eff}^{(m)}(t)$ as described in~\eqref{eq:H_eff}.
Formally, we can define $P_m$ to be the projector onto $\mathcal{H}_\text{eff}^{(m)}(t)$,
although we do not construct such a projection operator explicitly in memory.
\item Then evolve $\ket{\psi(t)}$ within this Hilbert space.
This can be formally written as $\tilde{U}(\delta t) = \exp(-i P_m H P_m \delta t/\hbar)$.
\item Finally, if necessary, truncate the resulting state.
This truncation can also be formally described as a projector $Q$,
but, once again, no such projection operator is explicitly constructed.
\end{enumerate}
Thus, the action of a full step of the algorithm can be formally described as the mapping
\[\ket{\psi(t)} \longmapsto Q \exp(-i P_m H P_m \delta t/ \hbar) \ket{\psi(t)},\]
where the essence of the method lies in determining the $m$-th order Hilbert space $\mathcal{H}_\text{eff}(t)$
(formally represented by $P_m$) and determining a suitable basis set to then truncate to (formally represented by $Q$).
We will use the same terminology to discuss error scaling behavior in section~\ref{sec:performance}.

One of the main advantages of the present method is that all of the aforementioned components lend themselves
very well to parallel execution on GPUs, enabling substantial speedups compared to sequential execution.
More precisely, the computation of $\exp(-i P_m H P_m \delta t/ \hbar)$ from the restricted Hamiltonian $P_m H P_m$
is highly parallelized, as it is performed via the repeated application of the Hamiltonian to the state (see App.~\ref{app:propagator}),
which can be done using existing highly efficient sparse matrix-vector multiplication (SpMV) algorithms.
Applying the truncation $Q$ is computationally trivial, but determining the set to be truncated
relies on a combination of sorting and searching.
The latter is implemented concurrently but is relatively diverging and therefore not entirely optimal.
Finally, determining $\mathcal{H}_\text{eff}(t)$ is, as mentioned above,
achieved using a matrix-free method which can be parallelized across every basis state,
but transferring the state $\ket{\psi(t)}$ from one effective Hilbert space to the next relies on
matching up different indices, which is also a branch-divergent problem and is likely currently the weak point of the method.

As such, the present algorithm features no local decompositions, in contrast to methods such as MPS or ML-MCTDH,
allowing the system to be tackled in its entirety all at once and enabling easier parallelization.
This, of course, is a double-edged sword: In pursuit of parallelization, we are required to manipulate the entire dataset
simultaneously, i.e., it becomes necessary to store the entire state vector, matrices etc.~in readily accessible RAM.
As GPU RAM sizes are currently often still an order of magnitude or more below what is available for CPUs,
this is the main practical computational limitation of the method,
though \enquote{unified memory} can alleviate this problem at the expense of becoming memory-bandwidth-limited.
We note that the parallelized and sparse nature of our method makes useful computations of time-complexity highly involved,
in part due to non-negligible I/O and memory access times relative to actual computational time~\cite{Bender2010}.

\section{Error analysis \& a benchmark system}\label{sec:performance}
In this section, we shall first investigate the origin and behavior of the errors induced by this method---%
that is to say, the deviations between the true time-evolved state and the time-evolved state as computed using \mname{}---%
and then benchmark the method by numerically investigating its internal convergence
and comparing results to previously published multiset-MPS results.

\subsection{General error analysis}\label{sec:error_analysis}
We first distinguish between the different possible sources of deviations from the true time evolution
and determine their relative importance using the terminology of $P_m$ and $Q$ that was established in section~\ref{sec:parallelization}.
The three principal sources of differences between the exact time evolution and the time evolution by the present method are:
\begin{enumerate}
\item The truncation of the time-evolution operator $U(\delta t) = \exp(-i P_m H P_m \delta t/\hbar)$ after a finite number of terms in its series expansion (discussed in detail in App.~\ref{app:propagator}),
\item The truncation of the state vector due to the projection $Q$, and
\item Artifacts due to the restriction into an effective Hilbert space with only $m$-th-order neighbors, as given by the projection $P_m$.
\end{enumerate}
The error due to the truncation of the series expansion of $U(\delta t)$ is essentially negligible,
as the expansion is terminated on a case-by-case basis using a convergence criterion.
This is enabled by its relatively \enquote{cheap} numerical nature.
Data demonstrating that this error is insignificant is shown in section~\ref{sec:benchmarking} and App.~\ref{app:convergence};
we refer the reader to references~\onlinecite{alhi10,alhi11} for detailed error analysis
and henceforth consider the error to be negligible compared to the other errors encountered in our method.

The error induced by the truncation $Q$, measured in terms of the wavefunction norm, is simply
$\norm\big{\ket{\psi} - Q\ket{\psi}}$.
The size of this error depends on the distribution of the expansion coefficients $\abs{\braket{n}{\psi}}^2$.
The exact nature of this distribution must depend on the Hamiltonian, the initial state and the basis that is chosen:
A general law predicting (at least asymptotically) the distribution of the expansion coefficients
would, arguably, constitute a rather profound statement on the foundations of quantum mechanics that we cannot provide here,
but at the same time, taming this distribution lies at the heart of \mname{}, so we will consider two different educated guesses:
an exponential distribution and a power law.

If we assume that the expansion coefficients in the basis $\mathfrak{B}$ decay exponentially,
i.e., $\abs{\braket{n}{\psi}}^2 \propto \alpha^{n}$ for some $\alpha \in (0,1)$, then if $\mathfrak{B}$ is countably infinite,
the truncation error due to projecting onto the $q$ highest-weight basis states is given by
\[\norm{\ket{\psi} - Q\ket{\psi}}
    = \sqrt{\sum_{k = q}^\infty \abs{c_k}^2}
    = \alpha^{q/2},\]
that is, if the coefficients decay exponentially, then the error induced by each truncation also decays
exponentially as a function of the number of basis states $q$ that the state is projected onto.
Naturally, this means that if the decay of the coefficients can be upper-bounded by some exponential function,
then the error features the same kind of exponential upper bound.
Because of this, an exponential distribution is a rather optimistic viewpoint on the possible distribution of coefficients.
Potential hand-waving justifications for an exponential distribution could be based on Lieb--Robinson-type arguments,
Anderson localization, or simply on the idea that an initially localized (in $\mathcal{H}$) state may spread exponentially
under a time-evolution operator $e^{-i H \delta t/\hbar}$, but none of these arguments are even close to rigorous.

On the other hand, a much more pessimistic but potentially also realistic distribution could be a power law:
$\abs{\braket{\psi}{n}}^2 \propto 1/n^{s}$ with $s \geq 1$---where $s = 1$ corresponds to Zipf's law,
which is, however, not normalizable if the Hilbert space is infinite-dimensional.
The truncation error in the case of a power law can be computed to be
\[\varepsilon(Q)
    \eqqcolon \norm{\ket{\psi} - Q\ket{\psi}}
    = \sqrt{\frac{1}{\zeta(s)} \sum_{k=q+1}^\infty \frac{1}{n^s}},\]
where $\zeta(s)$ is the Riemann zeta function which normalizes the coefficients.
We can bound this error by the standard Maclaurin--Cauchy integral technique to obtain:
\[\frac{1}{\sqrt{\zeta(s) (s-1) (q+1)^{s-1}}}
    \leq \varepsilon(Q)
    \leq \sqrt{\frac{\frac{1}{q+1} + \frac{1}{s-1}}{\zeta(s) (q+1)^{s-1}}}.\]
This means that, asymptotically, the error induced by the truncation $Q$ under a power law is
\begin{equation}
\varepsilon(Q) = \mathcal{O}\fleft(q^{-\frac{s-1}{2}}\fright).
\label{eq:epsilon_Q}
\end{equation}
As mentioned in section~\ref{sec:detruncation}, one should be note that the parameter $q$ contained in $Q$
is not exactly equal to the nominal truncation parameter \qnom{} due to the detruncation procedure.
The $q$ that occurs during the projection process $Q$ always satisfies $q_\text{nom} \leq q \leq q_\text{true}$, that is to say:
If all of the \qtrue{} basis states contained in the expanded effective Hilbert space are occupied by the wavefunction
prior to truncation, then $q = q_\text{true}$, whereas in the worst-case scenario where none
of the states added by expanding the basis via~\eqref{eq:H_eff} were occupied, $q = q_\text{nom}$.

Finally, we need to consider the error induced by the restriction to the subspace $\mathcal{H}_\text{eff}(t)$
as given in~\eqref{eq:H_eff}.
As the derivation of the bound for this error is somewhat involved, we relegate it to App.~\ref{app:restriction_error}.
We find that the error induced by this restriction can be upper-bounded as:
\begin{equation}
\varepsilon(P_m) \leq \frac{1}{(m+1)!} \left(\frac{\alpha \delta t}{\hbar}\right)^{m+1}.
\label{eq:epsilon_Pm}
\end{equation}
However, determining $\alpha$ (which is characterized in more detail in App.~\ref{app:restriction_error}) in practice is challenging.

An important question to consider is: How can one distinguish between these three sources of errors in a calculation?
The first two errors are relatively simple to detect:
An improperly truncated time-evolution operator will most likely not be unitary (see App.~\ref{app:Taylor}),
such that one can compare the norm and potentially other conserved quantities such as the total energy
before and after the application of the time-evolution operator to determine whether such an error is present.
Similarly, the truncation error due to $Q$ has an immediate effect on the norm of the wavefunction
and its effect is largely quantifiable by measuring the discarded weight.
The error due to the restriction to the effective Hilbert space via $P_m$, on the other hand,
is rather devious, as it will not affect the norm of the wavefunction.
An overly restrictive projection $P_m$ will essentially give rise to finite-size effects
such as reflections at \enquote{boundaries} in an abstract sense, which do not render the state manifestly unphysical,
but will instead effect seemingly good convergence to a wrong result.
However, as it is clear that increasing $m$ and increasing \qnom{} must both increase the fidelity of the results,
it is straightforward to check for convergence numerically.

In practice, it turns out that the truncation parameter \qnom{} is the most important convergence parameter and that,
for the most part, $m$ and $\delta t$ need to be above and below a certain threshold, respectively:
We demonstrate the crucial \qnom{} convergence in the next section~\ref{sec:benchmarking}
and demonstrate the less critical $m$ and $\delta t$ convergence in App.~\ref{app:convergence}.

\subsection{Benchmarking with a 1D Holstein system}\label{sec:benchmarking}
\begin{figure}
\centering
\includegraphics{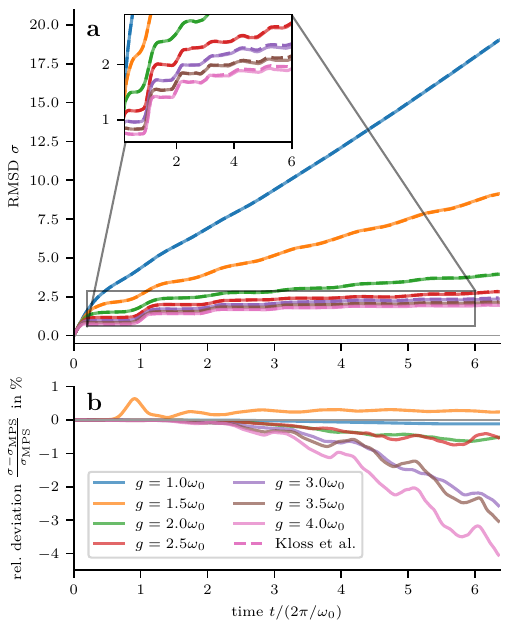}
\caption{A comparison between the RMSD values obtained using \mname{} and those of Kloss et al.~\cite{Kloss}
(compare to~Fig.~2 therein).
The initial state is a completely localized Franck-Condon-type excitation.
Full parameter list given in Tab.~\ref{tab:parameters_Kloss_1D}.
\textbf{a}: \mname{} (solid) and multiset-MPS~\cite{Kloss} results (dashed) for the RMSD~\eqref{eq:RMSD}.
\textbf{b}: Relative deviations between the RMSD of the two methods.}
\label{fig:benchmark}
\end{figure}
We benchmark \mname{} using the 1D non-equilibrium Holstein results given by Kloss et al.~in~\onlinecite{Kloss} as a reference,
which uses a single-site time-dependent variational principle (1-site TDVP) algorithm applied to a
multiset variant of MPS, in which several MPS are used simultaneously
in an effort to better deal with highly entangled states by separating them into less entangled individual MPS.

The main observable of interest in these calculations was the root-mean-square deviation (RMSD),
defined as the standard deviation of the excitonic\footnote{Reference~\onlinecite{Kloss} refers to electrons, but since we are considering only a single particle, this is, for the purposes of the Holstein Hamiltonian, equivalent to a single exciton.} 
position distribution function. The latter is given by
\[P\fleft(X_\text{exc} = x_n\fright) = \expval{\left(b^\text{exc}_{x_n}\right)^\dag b^\text{exc}_{x_n}} = \braket{\psi}{x_n}\braket{x_n}{\psi}.\]
Thus, the RMSD is defined as
\begin{equation}
\text{RMSD} 
\coloneqq \sqrt{\sum_{x_n=0}^{L-1} \left[x_n P\fleft(X_\text{exc} = x_n\fright) - \overline{X}_\text{exc} \right]^2 }, \label{eq:RMSD}
\end{equation}
with $\overline{X}_\text{exc}$ being the average exciton position
\[\overline{X}_\text{exc} = \sum_{x_n=0}^{L-1} x_n P\fleft(X_\text{exc} = x_n\fright) = \sum_{x_n=0}^{L-1} \braket{\psi}{x_n} x_n \braket{x_n}{\psi}.\]
The RMSD values obtained by \mname{} are compared to those computed by Kloss et~al.~in Fig.~\ref{fig:benchmark}.
The full list of parameters used in these calculations is given in Tab.~\ref{tab:parameters_Kloss_1D} of App.~\ref{app:parameters}.

There is generally good agreement between the results, but discrepancies increase with increasing time $t$
and increasing vibronic coupling strength $g$ up to a maximum relative deviation in RMSD of roughly \SI{4}{\percent}.
Intuitively speaking, this behavior is not particularly surprising, as the wavefunction spreads out as $t$ increases,
and larger $g$ causes higher-lying Fock states to become occupied,
increasing spreading in the vibrational Hilbert spaces and worsening truncation artifacts
while also increasing entanglement between the various vibrational and excitonic degrees of freedom.
As the $g = 4\omega_0$ case is the one with the largest discrepancies between the \mname{} and multiset-MPS results,
we consider its convergence behavior in more detail both in the following (for \qnom{})
and in App.~\ref{app:convergence} (for $m$ and $\delta t$).
\begin{figure}
\centering
\includegraphics{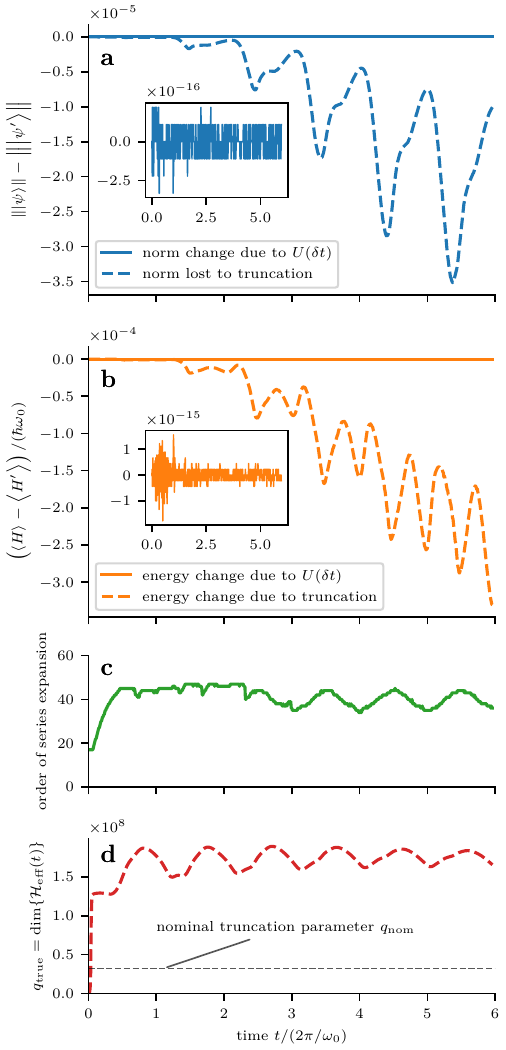}
\caption{Errors accumulated per timestep and diagnostic parameters from the $g=4\omega_0$ calculation of Fig.~\ref{fig:benchmark}.
See Tab.~\ref{tab:parameters_Kloss_1D} for the full list of parameters.
\textbf{a}: Norm loss \emph{per timestep} ($\delta t = 0.05/\omega_0$)
due to the finite-series expansion of $U(\delta t)$ and the truncation $Q$ of the state vector $\ket{\psi(t)}$.
Insets show a zoom on the $U(\delta t)$ truncation error, which is essentially zero due to its
convergence-based termination.
\textbf{b}: Same as \textbf{a}, but for the change in energy per timestep.
\textbf{c}: The order of the expansion of $U(\delta t)$ that was required for convergence (see App.~\ref{app:propagator} for details).
\textbf{d}: The dimension of the effective Hilbert space, i.e., \qtrue{} of~\eqref{eq:qtrue},
which expands and contracts depending on the structure of $\ket{\psi(t)}$.
This shows the dynamic and state-dependent relation between the fixed parameter \qnom{} and the actual dimension \qtrue{}.}
\label{fig:error_sources}
\end{figure}

We investigate the internal convergence of the results by analyzing
the \emph{norm} of the wavefunction and the \emph{total average energy} of the system,
both of which should be conserved over the course of an exact calculation
and allow for convergence checks in \mname{} without reference to an external calculation.
Changes in the norm can be indicative of wavefunction truncation $Q$ or of an ill-converged time-evolution operator---%
note that the truncation $Q$ may never increase the norm, but an ill-converged $U(\delta t)$ may both increase or decrease it.
Changes in the average total energy can also result from either of these,
although in this case, the truncation may also cause an increase.
Figure~\ref{fig:error_sources} illustrates the errors accumulated per timestep in these quantities,
verifying that the error attributable to the series expansion of $U(\delta t)$ is, as expected,
negligibly small due to its convergence-based termination.
On the other hand, the truncation error is not entirely negligible, indicating that the truncation parameter \qnom{}
is the main parameter controlling the convergence of the results.

\begin{figure*}
\centering
\includegraphics{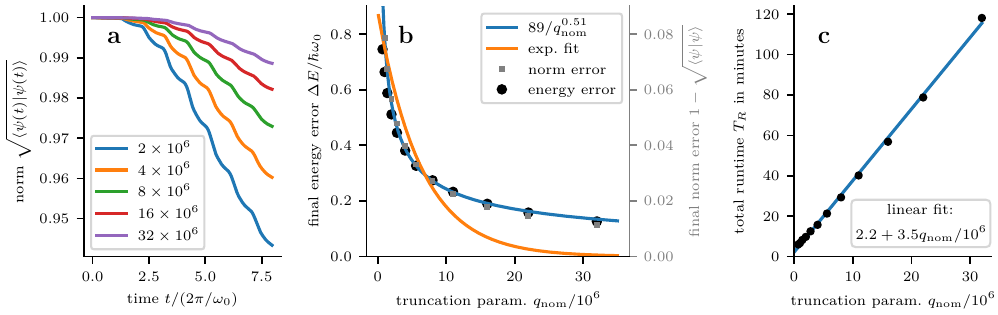}
\caption{Convergence as a function of the nominal truncation number \qnom{},
which is typically the most important parameter for convergence.
Note that \qnom{} is smaller than the true dimension \qtrue{}, as described in section~\ref{sec:detruncation}:
For the data shown here, $q_\text{true} \approx 5.03 q_\text{nom}$ across the various \qnom{} to within about \SI{1}{\percent}.
The norm $\sqrt{\braket{\psi}}$ and the average total energy $\expval{H}$ are used to gauge convergence here.
Apart from \qnom{}, data corresponds to the $g = 4\omega_0$ calculation of Fig.~\ref{fig:benchmark}.
Data shown in \textbf{b} and \textbf{c} is for the final time $t = 50/\omega_0$.
\textbf{a}: The time evolution of the norm for different values of \qnom{}, indicated in the legend.
\textbf{b}: The deviation from the exact value in terms of the total energy (black circles) and the norm (gray squares),
along with a least-squares fit of the norm error based on a power law (blue) or an exponential decay (orange).
\textbf{c}: The total runtime $T_R$ of the calculation on an Nvidia A100 (\SI{80}{\gibi\byte} HBM2e)
as a function of \qnom{} along with a linear least-squares fit.}
\label{fig:convergence_q}
\end{figure*}
Furthermore, we see in Fig.~\ref{fig:convergence_q} that, as expected, the errors decrease with increasing \qnom{}. 
The decay of the errors in the norm of the state seems to match a power law with
$\varepsilon(Q) \propto {q_\text{nom}}^{-0.51}$ quite well.
Then~\eqref{eq:epsilon_Q} suggests that in this situation---%
assuming we can extrapolate from the single-step error analysis performed above to a compounded error---%
the wavefunction coefficients may be distributed approximately as $\abs{\braket{n}{\psi}}^2 \propto n^{-2}$.
This power law is also indicated in Fig.~\ref{fig:coeff_distribution},
and we see that the true distribution matches this scaling reasonably well in the tail region for short to intermediate times,
but that the tail of the distribution further flattens out for later times.
Figure~\ref{fig:convergence_q} also reveals that the runtime scales remarkably linearly as a function of \qnom{}.

Finally, we remark that due to the high degree of parallelization, the present method is quite fast:
For the strong coupling case $g=4 \omega_0$, Kloss et al.~reported a total runtime of 156 hours
to reach $t_\text{max} = 40\omega_0$
(running as a single process on a Xeon E5-2690 v3 at \SI{2.60}{\GHz}),
whereas the present parallelized method required 94 minutes for the same calculation
on an Nvidia A100 (see App.~\ref{app:parameters} for further parameters of both methods).
However, a more thorough investigation of runtime comparisons across a wider range of hardware configurations,
code implementations and Hamiltonians would be required to allow for more general statements of relative
runtime-efficiencies.

\section{Application to vibronic nanoaggregates}\label{sec:aggregates}
To demonstrate the utility of the method, we apply it to small clusters of chromophores in one,
two and three dimensions by extending the Holstein Hamiltonian~\eqref{eq:holstein} to square or cubic lattices.
A central and basic quantity of interest in chromophore aggregates is the linear optical absorption spectrum.
In this section, we will use \mname{} to study two different influences on such optical spectra:
First, the effect of dimensionality on the optical properties of nanoaggregates, i.e.,
how the optical properties depend on whether the chromophores are arranged in a chain, a square grid or a simple cubic lattice.
Second, we investigate size-dependent effects of three-dimensional nanoaggregate cubes
by comparing optical spectra of cubic lattices with different side lengths.
To put the findings presented in this section into context,
a synopsis of the optical properties of vibronic dimers as a function of $g$ and $J$ is given in App.~\ref{app:dimers}.

In the dipole approximation and assuming classical light, the linear optical spectrum can be computed as~\cite{May2011Wiley,Schroeter2015PhysRep}
\[A(\omega) \propto \Re{\int_0^\infty \expval{\mu(t) \mu(0)} e^{i\omega t} \dd{t}},\]
where $\mu(t)$ is the system-dipole operator in the Heisenberg picture at time $t$
and the expectation value is taken with respect to the state of the system prior to its interaction with light.
To avoid degeneracies that might cause vibronic/vibrational and excitonic features in the spectra to overlap,
we choose non-commensurate parameters based on a Cy3 system~\cite{Kessing2022JPCL,Hart2021,Hart2022}:
$g = 0.71 \omega_0$ (corresponding to a Huang--Rhys factor~\cite{HuangRhys} of $S = 0.50$) and $J = 0.55 \omega_0$,
where the vibrational frequency is $\omega_0 / (2\pi) = \SI{34.5}{\THz}$ corresponding to $\tilde{\nu} = \SI{1150}{\per\cm}$---%
but we will largely continue to refer to $\omega_0$ as the base unit in the following.
Given these parameters and assuming the spectroscopy is performed at temperatures no greater than room temperature,
the initial state of the vibronic system is well-approximated by the global vacuum (i.e., zero-particle) state
$\ket{\Omega_\text{global}} = \ket{\Omega_\text{exc}} \otimes \ket{\Omega_\text{vib}}$.
For wavelengths of light much larger than the size of the aggregate itself,
the system-dipole operator in the Schrödinger picture can be expressed as
\[\mu \propto \sum_j \left[ \dyad{\Omega_\text{exc}}{j} + \dyad{j}{\Omega_\text{exc}} \right],\]
which acts diagonally on the vibrational Hilbert spaces.
This assumes that the chromophores have no static dipole moment and
that the transition dipole moments of all chromophores are aligned and not orthogonal to the polarization of the incoming light.

Then, under the aforementioned assumptions, the dipole operator $\mu$ acting on the initial state creates an equal superposition 
of Franck--Condon excitations at each chromophoric site in the aggregate:
\begin{equation}
\ket{\psi} = \mu \ket{\Omega_\text{global}} \propto \frac{1}{\sqrt{N}} \sum_{j} \ket{j} \otimes \ket{\Omega_\text{vib}},
\label{eq:init_state_spec}
\end{equation}
where $N$ is the number of sites.
Conversely, $\mu$ acting on any single-exciton state projects the excitonic part of the state back to $\ket{\Omega_\text{exc}}$
while preserving all of its vibrational degrees of freedom.
Therefore, we can use \mname{} to compute $\expval{\mu(t) \mu(0)}$ by evolving the state $\ket{\psi}$
up to time $t$ and then computing the overlap of its vibrational state with $\ket{\Omega_\text{vib}}$.
Taking the Fourier transform of the result then provides us with the typical frequency-domain linear absorption spectrum.

To avoid artifacts in the Fourier-transformed spectra arising from the hard signal cutoff at the final time of the calculation,
the computed signal $\expval{\mu(t) \mu(0)}$ is damped by an exponential function $e^{-t / \tau}$
prior to Fourier-transforming.
This procedure can be interpreted as an application of the quantum-jump method as described in App.~\ref{app:jumps}.
The lifetime $\tau$ is chosen to be as long as possible while still removing ringing artifacts,
but, to allow for fair comparisons between different systems, it is chosen to be identical for all results shown in this section.
The computed signal of J-aggregates decays much more slowly than that of the H-aggregates and is therefore the limiting factor
(see also Fig.~\ref{fig:rt_signal_4} in App.~\ref{app:convergence}), necessitating the use of a rather short $\tau = \SI{80}{\fs}$,
corresponding to lifetime-broadening with a full-width at half-maximum of $\SI{133}{\per\cm}$ or $0.115 \omega_0$ in the spectra.

Though it is mathematically straightforward to generalize the Holstein Hamiltonian~\eqref{eq:holstein}
to two and three dimensions by adding corresponding excitonic coupling terms,
we point out a physical caveat that occurs in three dimensions:
A simple cubic lattice of chromophores whose transition dipole moments are fully aligned
and that interact solely via dipole-dipole coupling
will not couple equally in all three dimensions due to the nature of dipole-dipole coupling
(though other types of coupling may be able to produce such isotropic coupling).
In other words, such a dipole-coupled cubic lattice would likely be of H-type along one or two axes
and of J-type along the remaining axes.
A similar issue may arise in two-dimensional J-aggregates.
Nevertheless, we present the two- and three-dimensional aggregates with fully isotropic couplings to avoid introducing
additional complexity and asymmetry to the problem, as it turns out that even this simple, symmetric
and disorder-free Hamiltonian gives rise to interesting non-trivial behavior.

\begin{figure*}
\centering
\includegraphics{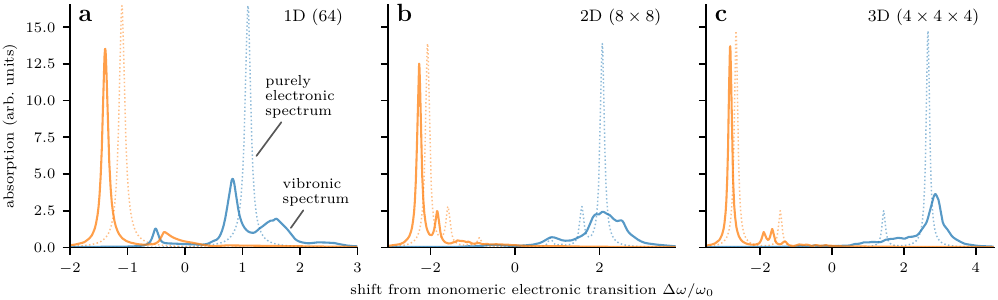}
\caption{The linear absorption spectra of 64 chromophores arranged in a chain (\textbf{a}),
a square grid (\textbf{b}) or a simple cubic lattice (\textbf{c}) under H- or J-type coupling.
Orange (red-shifted) lines represent J-aggregates, and blue (blue-shifted) lines show H-aggregates.
Solid lines show the vibronic spectra with $g = 0.71\omega_0$ ($S = 0.5$) as computed using \mname{},
while dotted lines are the corresponding purely electronic spectra ($g = 0$) for reference
(from exact diagonalization of the electronic system).
Full list of parameters is given in App.~\ref{app:parameters}.}
\label{fig:abs_spec_3}
\end{figure*}
The effect of dimensionality is investigated in Fig.~\ref{fig:abs_spec_3},
which shows a set of 64 emitters arranged in a chain of length 64, an $8 \times 8$ grid or a $4 \times 4 \times 4$ cubic lattice.
Of the three, the 1D systems exhibit the simplest spectra for both H- and J-coupling:
Both essentially resemble the corresponding vibronic dimers (cf.~Fig.~\ref{fig:Jseries})
with increased red- or blue-shifting, some broadening and splitting.
This similarity between the 64-site chains and dimers can likely be attributed
to the fact that the initial state after excitation~\eqref{eq:init_state_spec}
is an eigenstate of a purely electronic system in one dimension if the chain length $L$ is either $L = 2$ or $L \to \infty$.
A chain length of $L = 64$ is already sufficiently long to minimize finite-size effects~\cite{Spano2010ACR},
rendering the symmetrically excited state~\eqref{eq:init_state_spec}
nearly an eigenstate of the electronic part of the Hamiltonian.

By contrast, when the same number of emitters is packed into two or three dimensions,
the side length $L = N^{1/d}$ of the lattice becomes smaller, taking the system out of the large-$L$ regime
and giving rise to several electronic bright states,
as is confirmed by the appearance of further lines in the purely electronic spectra shown in Fig.~\ref{fig:abs_spec_3}.
For the J-aggregates in two and three dimensions, the effect of vibronic coupling
is largely confined to slight red shifts and vibronic splitting into relatively sharp lines
while remaining qualitatively similar to the corresponding electronic spectra.
On the other hand, the effect on H-aggregates is rather nontrivial 
and can be described as strong, somewhat unspecific broadening and reduction of the signal intensity.
Such stark differences in optical signatures between otherwise equivalent vibronic H- and J-aggregates
in the intermediate-coupling regime are in line with existing knowledge~\cite{Fulton1964JCP,Spano2010ACR,Hestand2018ChemRev},
but, though the impact of geometry on vibronic spectra and transport has been the subject of extensive research~\cite{Zhao2007JPCC,Spano2011JPCB,Eisfeld2011JLum,Chuang2016PRL,Chuang2019Chem,Giavazzi2025JCP},
the direct influence of 1D versus 2D or 3D arrangements on equivalent sets of emitters has not yet been explored in depth.

Since the total Hilbert space dimension of these systems is entirely identical and these three systems differ only in their
physical geometry (which, in turn, also influences the connectivity of the Hamiltonian), these systems are good testbeds
for examining the impact of geometry on \mname{}:
Increasing the dimension from 1D to 2D increased the runtime by a factor of two,
while increasing from 2D to 3D increased the runtime by roughly another factor of 2.5,
but the latter also required unified memory to run at the given convergence parameters, which significantly increases runtime
(see Apps.~\ref{app:parameters} and \ref{app:convergence} for details).

\begin{figure}
\centering
\includegraphics{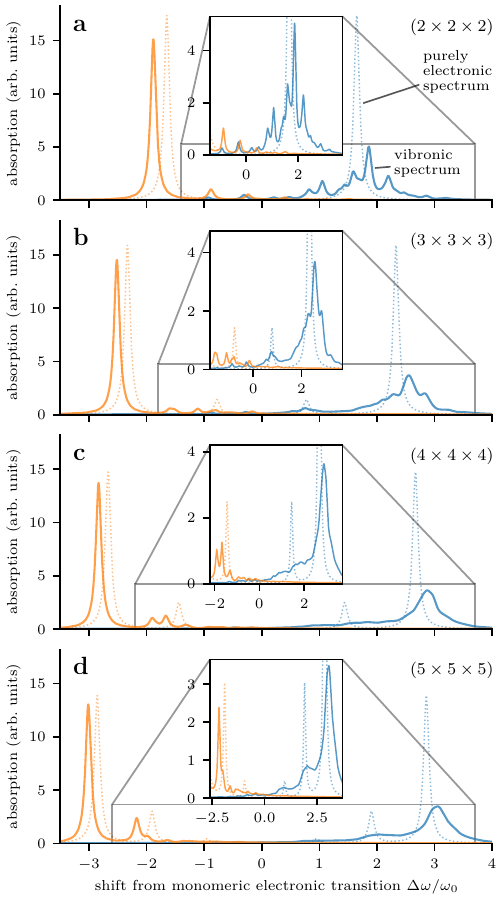}
\caption{The linear absorption spectra of three-dimensional simple cubic H and J nanoaggregates in four different sizes,
ranging from $2\times 2 \times 2$ in \textbf{a} up to $5\times 5 \times 5$ in \textbf{d}.
As in Fig.~\ref{fig:abs_spec_3}, orange lines represent J-aggregates, blue lines show H-aggregates;
solid lines are vibronic spectra computed with \mname{}
and dotted lines are purely electronic spectra from exact diagonalization.
Insets highlight the vibrational fine structure in the H-aggregates.
All spectra have been normalized with respect to their respective number of chromophores to make the signal intensity
comparable across system sizes.
Full list of parameters is given in Tab.~\ref{tab:parameters_abs_spec_3d} in App.~\ref{app:parameters}.}
\label{fig:abs_spec_4}
\end{figure}
The second effect, that of finite-size effects in 3D lattices, is investigated in Fig.~\ref{fig:abs_spec_4}.
Beginning with the largest system sizes, the $5 \times 5 \times 5$ H-aggregate
essentially resembles a broadened, slightly blue-shifted version of the corresponding purely electronic signal.
The equivalent J-aggregate reveals slightly more structure, but most nontrivial features are of very low intensity.
Though the $5 \times 5 \times 5$ spectra are, therefore, not particularly feature-rich,
they do reveal that this system size already proves to be remarkably self-decohering:
Despite the system Hamiltonian~\eqref{eq:holstein} containing only three distinct frequencies ($g$, $J$, $\omega_0$)---%
rendering each individual \enquote{phonon bath} monochromatic and degenerate across sites---%
we observe that the system size and geometry alone suffice to serve, in aggregate, as a decohering bath to the excitonic states.
Reducing the system size to $4 \times 4 \times 4$ and $3 \times 3 \times 3$ reveals sharper
and more structured features in the spectra of both the H- and J-aggregates, 
beginning to undo the self-decohering effects observed in the $5 \times 5 \times 5$ lattice.
However, arguably the most interesting behavior arises from the simple $2 \times 2 \times 2$ systems.
These systems are already electronically distinguished, as the optically excited state~\eqref{eq:init_state_spec}
is now once again an eigenstate of the electronic system,
unlike in the larger systems where the separation into edge and bulk states led to several electronic lines.
As such, the vibronic J-aggregate displays, for the most part, a very familiar-appearing vibronic progression.
However, despite the exceptionally simple purely electronic absorption spectrum of the $2 \times 2 \times 2$ systems,
the H-aggregate reveals a surprisingly rich and complex structure of distinct vibronic resonances.
These resonances vaguely resemble the familiar weight-redistributed vibronic H-dimer spectra
(cf.~Figs.~\ref{fig:Jseries} and \ref{fig:gseries_Hdimer}), but heavily split and shifted into an intricate pattern.
Just as with the self-decohering property described above,
this rich structure may be surprising in light of the highly symmetric and simple Hamiltonian that produced it;
we can verify that this rich structure is indeed due to the geometry by comparing it to the spectrum of an eight-site chain
shown in Fig.~\ref{fig:abs_spec_n8} in App.~\ref{app:n8}.

As \mname{} computes these linear spectra from non-equilibrium dynamics,
the data also provides direct access to other time-dependent observables.
For example, despite the striking differences in spectral properties shown above,
if the initial state is instead a fully localized Franck--Condon excitation (as was used in sec.~\ref{sec:benchmarking}),
then both the exciton and phonon dynamics are completely invariant under $J \mapsto -J$,
in agreement with previous findings in one-dimensional systems~\cite{Kessing2022JPCL}.
The evolution of the exciton distribution of such an initially localized excitation
in the $5\times 5 \times 5$ lattice is visualized in Fig.~\ref{fig:both_agg_n5x5x5} (Multimedia available online)
in App.~\ref{app:video}.
\begin{figure}
\centering
\includegraphics{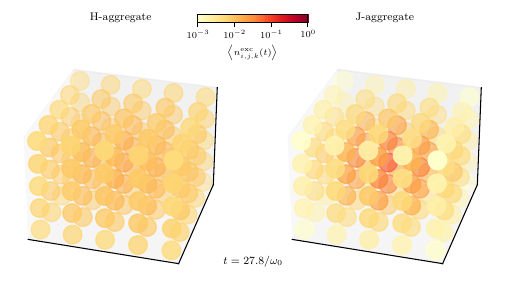}
\caption{(Multimedia available online/see the ancillary file \texttt{both\_agg\_5x5x5\_delocalized.mp4})
The evolution of the exciton distribution $\expval{n_{i,j,k}^\text{exc}(t)}$
following a symmetrically delocalized optical excitation~\eqref{eq:init_state_spec}
in H- or J-type $5 \times 5 \times 5$ cubic lattices.
Note that the color scale is logarithmic.
Data is taken from the $5 \times 5 \times 5$ calculations shown in Fig.~\ref{fig:abs_spec_4}.}
\label{fig:both_agg_n5x5x5_delocalized}
\end{figure}
On the other hand, if the initial state is the symmetrically delocalized initial state~\eqref{eq:init_state_spec},
then the excitonic position-space dynamics differ between the H- and J-systems.
Figure~\ref{fig:both_agg_n5x5x5_delocalized} (Multimedia available online) illustrates
this in the $5 \times 5 \times 5$ lattice:
The exciton undergoes \enquote{breathing} dynamics as it transitions from a symmetrically delocalized state
to a state that is mostly localized on the central site of the lattice and then back.
However, in the case of H-type coupling, these dynamics quickly die down and dephase,
while the J-type system repeats this process nearly indefinitely.
This can be understood as a visualization of coherent exciton dynamics in the case of the J-system,
in which the optical initial state~\eqref{eq:init_state_spec}, being a superposition of a set of low-quasimomentum states,
coherently oscillates between the different sites as a result of the well-defined phases between its quasimomentum states,
while the initial state in the H-aggregate (which is concentrated near the top of the excitonic band)
quickly explores the vibrational degrees of freedom and then rapidly self-decoheres.
This interpretation is further supported by the real-time signal of $\expval{\mu(t) \mu(0)}$
(Fig.~\ref{fig:rt_signal_4} in App.~\ref{app:convergence}),
whose Fourier transform was used for Fig.~\ref{fig:abs_spec_4};
said real-time signal also provides further insight into the underlying decay rates prior to the lifetime-damping procedure.

With regard to the performance of \mname{} in these 3D systems, one observes that the method,
naturally, performs better for smaller systems (see Tab.~\ref{tab:parameters_abs_spec_3d} for details).
In particular, the total Hilbert space dimension of the $2 \times 2 \times 2$ lattice that was considered here
is approximately \num{69e9}, which is only three orders of magnitude larger than the
$q_\text{true} \approx \num{32e6}$ that \mname{} employed (for $m = 2$, $q_\text{nom} = \num{16e6}$)
and therefore allowed for essentially untruncated results---for example, the final error in the norm here was less than $\num{3e-14}$.
On the other hand, the total Hilbert space dimension of the $5 \times 5 \times 5$ lattice shown here is over $10^{152}$,
which, combined with the relatively higher connectivity in three dimensions,
put much more pressure on the truncation procedure employed by \mname{}.
Interestingly, the data shows that the J-aggregate is much better suited to a truncated-vector representation than the H-aggregate,
as the former lost less than \SI{0.8}{\percent} of its norm at the end of the calculation while the latter lost nearly
\SI{70}{\percent} of its norm in the same time. This is likely due to the relationship between the initial state
and the eigenstates of the system and the induced self-decoherence discussed above.
We highlight that, despite this decay in norm, the spectra shown in Fig.~\ref{fig:abs_spec_4} are still well-converged
(demonstrated in Fig.~\ref{fig:conv_n5x5x5_abs_spec}), as the H-aggregate signal $\expval{\mu(t) \mu(0}$ itself
decays very rapidly, placing less emphasis on state fidelity after long simulation times.

As mentioned above, the results presented in this section assumed a very simple, highly symmetric coupling Hamiltonian.
Extensions to disordered systems in all parameters (vibrational frequency, local exciton energy, vibronic coupling)
as well as more realistic, anisotropic inter-site coupling $J$ are straightforward,
but would need to constitute the topic of future work.
More generally, since \mname{} computes non-equilibrium quantum dynamics on the state level,
extensions to other non-equilibrium quantities such as non-linear optical signals could also be considered in future work.

\section{Conclusions \& Outlook} \label{sec:conclusion}
We have presented a parallelized, efficient method for computing quantum dynamics of pure states
based on physically motivated co-evolving subspaces.
The concept is applicable to any system in which the basis admits a sparse representation of both
the Hamiltonian and the initial state, and it will be effective if
the state remains relatively sparse in the Hilbert space, which depends in part
on the \enquote{connectivity} of the Hamiltonian in the given basis.
As such, an appropriate basis can be identified as a crucial element to the method, whereas high dimensionality 
(both geometrically and in terms of the Hilbert space) or entanglement \textit{per se} are less central to its applicability.
The method compares favorably with existing state-of-the-art multiset-MPS calculations
of a 1D Holstein model benchmark.
Furthermore, we have demonstrated its applicability to higher-dimensional geometries by using it to
compute linear optical spectra of model nanoaggregates in one, two and three dimensions.

We would like to highlight three interesting possible extensions of the method:
First, the method is readily extensible to time-dependent Hamiltonians that satisfy the above criteria,
as long as Trotterization along the time axis is an acceptable approximation.
In other words, if a time-dependent Hamiltonian can be well-approximated as a piecewise constant-in-time Hamiltonian,
then the method trivially extends to such a system by simply adjusting the Hamiltonian accordingly at each timestep
(more specifically: by passing time as an argument to the matrix-free function that yields the action of the Hamiltonian).
Particularly well-suited time-dependent Hamiltonians include those that exhibit changes only in the magnitude of their coupling
and not in their graph structure, e.g., time-dependent driving of an otherwise constant operator such as $\cos(\omega t)\sigma_x$,
etc.---a simple albeit ubiquitous form of time-dependent Hamiltonian.

Second, the concepts described in this publication can be extended to the Heisenberg picture,
that is, to the time evolution of operators rather than state vectors.
When applied in such a manner, the method would likely overlap conceptually with
the recently developed and highly efficient Pauli/Majorana Propagation methods~\cite{Rall2019PRA,Begusic2025PRXQuantum,Miller2025arxiv,Rudolph2025arxiv},
as well as, to a certain extent, kernel polynomial methods~\cite{Weisse2006RMP}.
Such a modified Heisenberg-based approach could prove to be beneficial over the Schrödinger-state-based method described here
if it is known that only a certain observable will be of interest---%
or if the operator that is being propagated is a density operator, leading us to the next and final point.

Finally, the method can conceivably be extended to open quantum systems dynamics:
In the simplest case, by vectorizing a master equation, or by employing a Heisenberg-type evolution
that could allow for direct propagation of density matrices,
and then effectively replacing the Hamiltonian in its role as the dynamical generator with the Lindblad or Redfield superoperator.
This provides direct access to the perturbative system--bath coupling regime~\cite{Fruchtman2016SciRep},
but can also be combined with existing techniques such as pseudomodes~\cite{Garraway1997PRA,Tamascelli2018PRL}
to access the non-Markovian, non-perturbative regime in a controlled manner.
Furthermore, using techniques such as quantum jumps~\cite{Dalibard1992PRL,Plenio1998RMP}
(a very simple example of which was already demonstrated in the present publication)
or chain mapping~\cite{Prior2010PRL,Chin2010JMP}
and T-TEDOPA~\cite{Tamascelli2019PRL,LeDe2024JCP,Dunnet2021JCP,Lambertson2024JCP},
which transform non-Markovian dissipative environments into chains of coupled harmonic oscillators,
even the current wavefunction-based approach may be amenable to treating open quantum systems, which could then,
for example, be further extended using the transfer-tensor method~\cite{Cerrillo2014PRL,Rosenbach2016NJP}.

\begin{acknowledgments}
I am indebted to Martin Plenio and Susana Huelga for their support, supervision and many valuable discussions,
and to Salvatore Manmana for his tireless enthusiasm, encouragement and truly in-depth discussions on this project.
I am especially grateful to Jianshu Cao for hosting me in his group and for the inspiring and supportive environment
he provided during the development of this work.
I also thank the many colleagues who provided feedback and insightful discussions on this manuscript and its underlying ideas,
in particular Koenraad Audenaert, Dario Tamascelli, Gabriela Wójtowicz, Thibaut Lacroix,
Carlos Munuera Javaloy and Nicola Lorenzoni.
Furthermore, I would like to thank Benedikt Kloss for providing access to the data used in Fig.~\ref{fig:benchmark},
and to Monodeep Chakraborty for pointing out relevant literature that extended the LFS method.

I acknowledge generous scholarships and travel funds provided by the \textit{Studienstiftung des deutschen Volkes},
as well as funding provided by:
the German Research Foundation (DFG) -- 217133147/SFB 1073, project B03,
the German Federal Ministry of Science (BMFTR) under the project SPINNING (Grant No.~13NI6215),
and the BMFTR project PhoQuant (Grant No.~13N16110).
I also gratefully acknowledge computational resources provided by the state of Baden-W\"urttemberg through bwHPC
and the DFG through Grant No.~INST~40/575-1~FUGG (JUSTUS~2 cluster)
and GPU resources provided by the Institute of Theoretical Physics in G\"ottingen
and financed by the DFG and the Bundesministerium f\"ur Bildung und Foschung (BMBF).
\end{acknowledgments}

\section*{Author declarations}
The author has no conflicts to disclose.

\section*{Data availability}
The data that support the findings of this study are available from the corresponding author upon reasonable request.
The code that was used to generate the data presented in this study
is publicly available on GitHub: \url{https://github.com/rkevk/paces}.

\bibliography{article.bib}

\appendix

\section{Lossless bit-level compression of state information}\label{app:compression}
As an inherently sparse method, \mname{} needs to store information specifying to which basis state each
expansion coefficient in the vector belongs.
In this appendix, we sketch the lossless compression algorithm that is used to store this information more efficiently.
This compression is unrelated to the notion of effective Hilbert spaces;
as such, this section will be phrased independently of that language to avoid confusion.

As the method is designed for many-body systems, the underlying Hilbert space $\mathcal{H}_\text{tot}$
will generally be given as a tensor product of smaller Hilbert spaces,
$\mathcal{H}_\text{tot} = \bigotimes_{i=1}^L \mathcal{H}_i$.
For simplicity, assume that all local Hilbert spaces $\mathcal{H}_i$ are identical and have the same dimension,
such that $\mathcal{H}_\text{tot} = \mathcal{H}_i^{\otimes L}$:
Generalizing to varying dimensions is straightforward, but clutters the explanation.
Furthermore, let each local Hilbert space be represented in some enumerated basis $\{\ket{1}, \ket{2}, \ldots, \ket{d}\}$
such that each state in the total Hilbert space can be uniquely identified as
$\ket{n_1, n_2, \ldots, n_L}$ with $1 \leq n_i \leq d$ for all $1 \leq i \leq L$.

The arguably simplest way of storing which coefficient belongs to which basis state
is to save the label of the basis state in a separate array, e.g.,
the basis states belonging to the entries in the vector
$\begin{pmatrix} c_1\\ c_2\\ c_3\end{pmatrix}$ could be saved in an $3 \times L$ array
\[A = \begin{bmatrix}
    n_1^{(1)} & n_2^{(1)} & \cdots & n_L^{(1)}\\
    n_1^{(2)} & n_2^{(2)} & \cdots & n_L^{(2)}\\
    n_1^{(3)} & n_2^{(3)} & \cdots & n_L^{(3)}
\end{bmatrix}.\]
A considerable issue with this approach lies in the fact that modern computing hardware
will store such arrays with at least one byte (8 bits) per element,
and memory access patterns often even put single-byte arrays at a relative disadvantage compared to
arrays whose \emph{words} (the fundamental unit stored in an array) are 2 or 4 bytes long (16 and 32 bits, respectively).
However, the local Hilbert-space dimensions of a many-body problem are almost always much smaller than the $2^8 = 256$
or $2^{32} = \num{4294967296}$ that one and four bytes provide, respectively.
What this means in practice is that some or most of the bits in the array $A$ will always be zero and thus wasted.
For example, if the local Hilbert space dimension is $16 = 2^4$, then $A$, if stored as an array of 1-byte words,
will contain twice as many bits as necessary, and if it is stored as an array of 4-byte words,
then the first 28 bits of every single element of the matrix are guaranteed to be wasted memory space.

Since changing the local dimension of the physical system is not a generally feasible strategy
and computing hardware does not normally allow for words of arbitrary bit-lengths (especially not when mixed in a single array),
the approach that is used in the current implementation of \mname{} is to store the information
in a compressed form which simply deletes the always-zero bits using bit-shift operations.
For example, assuming that each local Hilbert space has a dimension of $d = 16 = 2^4$,
a 32-bit (= 4-byte) word can losslessly store the index of up to eight sites.
Instead of $A$, the lookup table is then stored as an array
\[C = \begin{bmatrix}
    w_1^{(1)} & w_2^{(1)} & \cdots & w_\Omega^{(1)}\\
    w_1^{(2)} & w_2^{(2)} & \cdots & w_\Omega^{(2)}\\
    w_1^{(3)} & w_2^{(3)} & \cdots & w_\Omega^{(3)}
\end{bmatrix}\]
where $\Omega = \left\lceil\frac{L \log_2{d}}{\text{wordsize}}\right\rceil$
with the wordsize given in bits.
The first element of the $r$-th row in $C$ contains the bits (assuming $d = 16$ and a wordsize of 8 bits for brevity)
\[w_1^{(r)} = \left(
    b_{11} b_{12} b_{13} b_{14} \
    b_{21} b_{22} b_{23} b_{24}
    \right),\]
where each $b_{jk}$ represents a single bit with $j$ the physical site index in the Hilbert space
and $k$ the index of the bit within that site.
The second element $w_2^{(r)}$ of the $r$-th row would then first contain the four bits representing the third physical site,
then the fourth physical site, etc.

As a practical example, let the underlying Hilbert space consist of $L = 8$ physical sites each with a local dimension $d = 16 = 2^4$.
Suppose the current state of the wavefunction is
\[\frac{3}{5} \ket{6, 0, 0, 1, 0, 2, 0, 15} + \frac{4}{5} \ket{14, 0, 0, 0, 0, 0, 0, 1}.\]
Then the vector that \mname{} stores is simply
$\begin{pmatrix} 0.6 \\ 0.8 \end{pmatrix}$,
and the corresponding \emph{uncompressed} lookup table $A_2$ would be 
\[A_2 = \begin{bmatrix}
    6 & 0 & 0 & 1 
        & 0 & 2 & 0 & 15\\
    14 & 0 & 0 & 0
        & 0 & 0 & 0 & 1
\end{bmatrix},\]
We will now demonstrate how $A_2$ is brought into the compressed form, assuming 16-bit words.
Consider the first row of $A_2$:
The bit-level representation of the first \enquote{occupation number}, 6, in a 16-bit word is
\[0000 0000 0000 \underline{0110},\]
where only the last four, underlined bits can ever be non-zero due to the dimension of the local Hilbert spaces.
What the compression algorithm must now do is take those last four bits of every word and left-shift
them by an appropriate amount (in this case: 12 bits) so as to leave no empty bits.
This is repeated for all eight physical sites and both rows.
The result of compressing $A_2$ is then:
\[ \begin{bmatrix}
\left( 
    0110 \ 0000 \ 0000 \ 0001
    \right) & 
\left( 
    0000 \ 0010 \ 0000 \ 1111
    \right) \\
\left( 
    1110 \ 0000 \ 0000 \ 0000
    \right) &
\left( 
    0000 \ 0000 \ 0000 \ 0001
    \right)
\end{bmatrix},\]
where each row now spans two 16-bit words.
In general, the bit-lengths of the individual sites may not be commensurate with the length of the word:
Some care must then be taken when bits belonging to a single site cross an inter-word boundary.
Note that we need not concern ourselves with the endianness of the architecture
as long as the programming language we use supports bit-shifts in multi-byte integers
in a natural way that emulates all bits being contained in a single \enquote{unit}.
One may note that this compression scheme is similar to that employed in Pauli Propagation~\cite{Rudolph2025arxiv},
but generalized to arbitrary and varying local dimensions.

\mname{} takes an uncompressed standard-number array of the form $A$ when reading the initial vector
as input from the user, but then immediately compresses this array into the $C$ form and never fully decompresses it.
Naturally, this means that some parts of the algorithm must be written in a way that acts directly on the
lookup tables without decompressing them in their entirety, which adds some implementation and runtime overhead,
but the savings in memory far outweigh the downsides here.
For example, if there are \num[print-unity-mantissa=false]{1e7} basis states in the current Hilbert space
with an underlying tensor-product space of 100 sites with a local dimension of 8 per site,
then the compressed format ($C$) using 4-byte words requires \SI{382}{\mebi\byte} to save the lookup table
while the uncompressed format ($A$) requires \SI{3.73}{\gibi\byte} under the same conditions.
Since these lookup tables are very heavily copied and manipulated during certain steps of the algorithm,
such a drastic increase in memory requirements would be unacceptable while the concomitant increase in runtime is not.

It should be noticed that, though this scheme is much more memory-efficient than simply storing the basis state information
without compressing away empty bits, it is not necessarily the best compression scheme for this situation.
For example, hash tables might provide an alternative and potentially even more memory-efficient solution to the problem.
Exploring such avenues would need to be the matter of future research.

\section{Sketch of matrix-product state methods}\label{app:mps}
A family of very widespread and highly successful methods in modern computational physics are matrix-product state (MPS) approaches,
which built on earlier density-matrix renormalization group (DMRG)~\cite{White1992,WhiteNoack1992,White1993}
and numerical renormalization group concepts~\cite{Wilson1975}.
We briefly present the core ideas behind these methods here, with specific regard to how they compare to the present method.
See, e.g., refs.~\onlinecite{Bridgeman,Schollwoeck,Paeckel2019} for more detailed modern reviews of these methods.

MPS methods deal with the curse of dimensionality by locally decomposing and compressing vectors and operators:
In a $d^L$-dimensional Hilbert space consisting of $L$ sites (each with a local dimension $d$),
an arbitrary state vector $\ket{\psi}$
\begin{equation*}
\ket{\psi} = \sum_{\sigma_1, \ldots, \sigma_j} \psi_{\sigma_1,\ldots,\sigma_L} \ket{\sigma_1, \ldots,\sigma_L},
\end{equation*}
where $\psi_{\sigma_1,\ldots,\sigma_L}$ are the typical state vector coefficients, can be rewritten as
\[\ket{\psi} = \sum_{\sigma_1, \ldots, \sigma_j} A^{(1)}_{\sigma_1} \cdots A^{(L)}_{\sigma_L} \ket{\sigma_1, \ldots,\sigma_L}\]
by repeatedly reshaping $\psi_{\sigma_1,\ldots,\sigma_L}$ and applying Schmidt decompositions to the resulting matrices.
Each of the $A^{(i)}_{\sigma_i}$ in this equation is a matrix, and since there are $d$ matrices associated to each site
(each matrix corresponding to a different local basis state $\sigma_i$),
each $A^{(i)}$ constitutes a third-order tensor.
Thus, MPS decomposes the expansion coefficient $\psi_{\sigma_1,\ldots,\sigma_L}$,
which can be regarded as a order-$L$ tensor, into a set of $L$ third-order tensors.

The two main merits of the MPS approach are that there is now a physically motivated local decomposition which may ease calculations,
and that there is a natural way to compress such a decomposed state by applying singular value decompositions (SVD)
to each $A^{(i)}$ and truncating the lowest-weight singular values such that only $\chi$ singular values are retained.
One usually specifies the maximum value $\chi$ that is employed along the MPS, $\chi_\text{max}$,
which is then called the (maximum) \emph{bond dimension}.
The truncated $A^{(i)}_{\sigma_i}$ are then matrices of dimensions less than or equal to $\chi_\text{max} \times \chi_\text{max}$.
Physically, this truncation of the matrices corresponds to truncating the entanglement between lattice sites.
Because of this, combined with the area law of entanglement, certain geometrically low-dimensional Hamiltonians are known to admit
highly efficient MPS representations~\cite{Verstraete2006}.
However, general systems, especially highly interconnected or (geometrically) higher-dimensional ones,
generally do not exhibit such \enquote{nice} properties and may suffer significantly from the entanglement-based truncation:
Intuitively, we see that the MPS decomposition given above requires a natural one-dimensional ordering,
since any given tensor $A^{(i)}$ must contract only with its neighboring tensors $A^{(i-1)}$ and $A^{(i+1)}$;
thus, if a system does not exhibit such a 1D order, apparent long-range correlations and entanglement
may arise from forcing a 1D ordering onto the system, inhibiting the compression and complicating the computation of such a system~\cite{Lacroix2025,Lacroix2025arXiv}.
The method described in this publication, on the other hand,
does not impose such a 1D ordering on the system and is therefore geometry-agnostic,
eliminating the need for optimized \enquote{topologies} of local sites.

As described above, the standard SVD-based compression used for MPS effectively truncates the inter-site entanglement,
but it does not affect the---potentially very large---local dimension $d$ of the state being described by an MPS.
Therefore, further techniques have been developed to efficiently truncate the local dimension $d$ in MPS,
in particular, the procedure that later became known as local basis optimization (LBO)~\cite{LBO1,LBO2,Brockt,Schroeder}.
The basic idea behind LBO is to diagonalize the reduced density matrix at each site and then retain only those
eigenvectors of the reduced density matrix that correspond to the largest eigenvalues based on some cutoff weight,
similar to the truncation procedure used in our method (see App.~\ref{app:truncation}).

Another very important development in MPS methods has been the application of the Dirac--Frenkel
time-dependent variational principle (TDVP) to time evolution~\cite{Haegeman2011PRL,Haegeman2016PRB}.
The single-site variant of this method (see, e.g., ref.~\onlinecite{Paeckel2019} for details)
has several important advantages such as conservation of norm and energy and its applicability to a wide range of Hamiltonians.
An important characteristic of said single-site TDVP is that it operates in a manifold of MPS with fixed bond dimensions
(thus, in a sense, fixed entanglement), which provides boosts to numerical efficiency, but comes at the cost of potentially
misrepresenting the time-evolved state if the true time-evolved state becomes significantly more entangled over time
such that it cannot be accurately represented in the low-bond-dimension manifold that may have been sufficient at earlier times.
However, recent approaches such as ancillary Krylov spaces~\cite{Yang2020PRB},
adaptive one-site TDVP~\cite{Dunnet2021JCP} and controlled bond expansion~\cite{Gleis2023PRL}
have allowed single-site TDVP to successfully overcome this limitation.
These extensions provide single-site TDVP with automatic and adaptive schemes to increase local entanglement as required,
similar in spirit to the present method and, in particular, greatly increasing the utility of MPS with TDVP with regard
to bosonic subspaces and open quantum systems~\cite{Dunnet2021JCP}.

A core feature of all MPS methods is their local decomposition (and efficient truncation)
of an exponentially large wavefunction, which allows for simpler local manipulations on the individual matrices.
This decomposition often confers considerable benefits,
though it also brings with it drawbacks in the form of sequential system sweeps and SVD or QR decompositions,
which present not insignificant obstacles for full parallelization of the methods,
though there have recently been considerable efforts towards efficient GPU or hybrid CPU/GPU implementations
of MPS methods (e.g.,~\onlinecite{Li2020,Huang2021NatCompSci,Lyakh2022FAMS,Feng2022PRL,Lambertson2024JCP,vanDamme2024SciPost,Menczer2025JCTC}).
This stands in contrast to the fundamentally parallel nature of \mname{}, as discussed in section~\ref{sec:parallelization}---%
though, as noted there, the lack of local decompositions of the present method is also not without its drawbacks.

\section{Sparse matrices and their exponentials}\label{app:sparse}
In section~\ref{sec:sparseHam}, we pointed out that many common many-body Hamiltonians are naturally representable as sparse matrices.
We shall first demonstrate this sparseness using two example systems and then investigate a caveat that arises when
using sparse matrices as a generator for time evolution.

First consider the density of the single-exciton 1D Holstein model Hamiltonian as given in~\eqref{eq:holstein},
with a fixed truncated phononic dimension $d_\text{pho}$ and a chain length $L$.
The total Hilbert space dimension of such a model is $\dim{\mathcal{H}} = L \left(d_\text{pho}\right)^L$
and therefore a dense representation of the corresponding Hamiltonian will
contain a total of $L^2 \left(d_\text{pho}\right)^{2L}$ matrix elements.
For a small system of $L=9$ and $d_\text{pho} = 5$, 
assuming that each entry is saved as a 64-bit number, a dense matrix acting on this space would already
require over \SI{2248}{\tebi\byte} of memory to represent, clearly out of reach of current computing power.
However, the density of such a Holstein Hamiltonian can be computed to be
\[\frac{1}{L^2 d_\text{pho}^{L}} \left[ \left( 5L - 2 \right) - \frac{2L}{d_\text{pho}} \right],\] 
which decreases exponentially with chain length.
Plugging in the same $L$ and $d_\text{pho}$, we find that the non-zero matrix elements require only
\SI{587}{\mebi\byte} of memory, less than $\frac{1}{\num{4000000}}$ of the dense matrix, as mentioned in the main text.
By contrast, the single-particle TB Hamiltonian~\eqref{eq:tb} has a density of $(3L - 2)/L^2$:
Though the density still decreases substantially as a function of $L$,
the lack of exponential decay here is due to the TB model's single-particle nature.

As another example of a many-body example that is unrelated to the Holstein Hamiltonian,
consider a geometrically $d$-dimensional lattice of spin-$\frac{1}{2}$ with a side length $L$,
leading to a total Hilbert-space dimension of $2^{L^d}$.
If we choose the basis to be the tensor product of the local $\sigma_z$ basis,
then a nearest-neighbor (NN) interaction Hamiltonian of the type
\[\sum_{\substack{j, k\\ j, k \text{ NN}}} v_{j,k} \sigma_x^{(j)} \sigma_x^{(k)}\]
consists of $4d(L-1)L^{d-1} 2^{L^d - 2}$ matrix elements.
A Hamiltonian that combines this with $2^{L^d}$ diagonal terms thus has a density of
\[\frac{1 + d(L-1)L^{d-1}}{2^{L^d}},\]
indicating that its density also rapidly decreases as either of $L$ or $d$ increases.

\subsection{The action of the exponential of a sparse matrix} \label{app:propagator}
Having demonstrated that many-body Hamiltonians can often be given as sparse matrices,
we turn to a crucial issue that arises when applying these sparse matrices:
The time evolution operator~\eqref{eq:time_evolution}, $U(\delta t) = e^{-i H\delta t/\hbar}$ is, in general, not sparse. 
Intuitively, one may easily see this by considering the following:
Though the Hamiltonian matrix may be sparse due to the local nature of the interactions,
$U(\delta t)$ consists of all powers of $H$ and therefore contains many iterations and combinations of all local interactions,
such that it generally lacks this property of locality.
In other words, if two states with indices $j$ and $k$ are not strictly separated for symmetry reasons,
then the time evolution operator matrix element $U_{jk}(\delta t)$ between those two states is also non-zero.
As a result, although this transition probability may be infinitesimally small for almost all pairs $(j,k)$,
the propagator matrix will nevertheless not be representable as a sparse matrix without further approximation
(unlike the Hamiltonian, which is often sparse without any approximation).
Though it may seem as if all of the above analysis on sparseness is thus wasted, such is not the case.

The remedy lies in not explicitly calculating $U(\delta t)$ and then letting this operator act on the state vector,
but rather in applying each iteration of $H$ to the state vector individually and calculating the running sum.
More precisely, the current implementation uses either the scaling-and-squaring method for $e^A B$ described by
Higham and Al-Mohi~\cite{alhi11}, or a simplified Taylor-series-based variant thereof;
that is, instead of calculating
\[U(\delta t) = \sum_{n=0}^N \frac{\left(-i \delta t H/\hbar\right)^n}{n!},\]
which would be a dense matrix that could then be applied to $\ket{\psi}$,
we instead immediately apply the action of each $\frac{-i \delta t/\hbar}{n} H$ to the vector and keep only the resulting vector.
In \texttt{python}/\texttt{numpy} syntax, the core of the idea is:
\begin{lstlisting}[language=Python,
    basicstyle=\scriptsize\ubuntumono,
    backgroundcolor=\color{backcolour},   
    commentstyle=\color{codegray},
    keywordstyle=\color{magenta},
    numberstyle=\color{codegreen},
    stringstyle=\color{codepurple},
    showstringspaces=false]
def single_step_evolution(dt, input_vec, H, num_iter):
    current_vec = input_vec
    result      = input_vec
    for n in range(1, num_iter+1):
        current_vec = H.dot(current_vec)
        current_vec *= -1j * dt/(hbar * n) 
        result      += current_vec
    return result
\end{lstlisting}
This method of calculation only requires keeping track of one additional vector apart from the input and output vector,
instead of needing to save an entire potentially dense matrix.
Thus, we can evaluate such a time evolution very memory-efficiently using existing algorithms for SpMV operations~\cite{Steinberger}.
This is also an advantage over the Lanczos algorithm, which requires storing multiple such vectors,
as detailed in App.~\ref{app:Lanczos}.

\subsection{An effective Hilbert space up to neighbor order \texorpdfstring{$m = 2$}{m = 2} is not equivalent to a second-order Taylor expansion}\label{app:Taylor}
\begin{figure}
\centering
\includegraphics{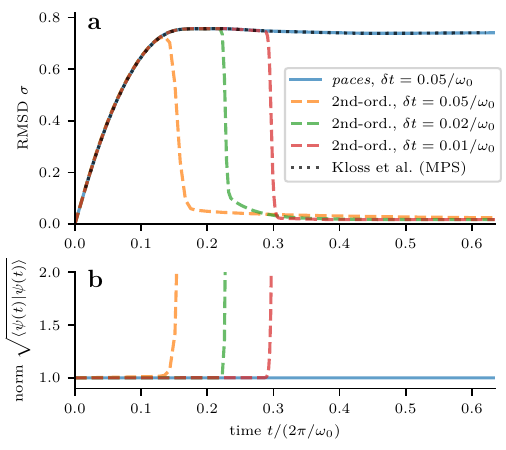}
\caption{The $g = 4\omega_0$ data shown in Fig.~\ref{fig:benchmark} produced using \mname{} with a maximal neighbor order $m = 2$
compared to data obtained using a second-order expansion of the time-evolution operator $U(\delta t)$.
The timestep used for \mname{} was $\delta t = 0.05/\omega_0$:
If the second-order expansion is used with a reduced timestep, its range of validity is slightly increased,
but it nevertheless fails to provide useful results after a relatively short time.
\textbf{a}: RMSD results of the different approaches, renormalized to take into account changes in norm.
\textbf{b}: The raw norm without renormalization, revealing a clear correlation between the onset of
unphysical increases in the norm and incorrect behavior of observables in the second-order truncation results.
Multiset-MPS data of Kloss et al.~\cite{Kloss} is also shown for comparison (as in Fig.~\ref{fig:benchmark}).
Apart from the restriction of the series expansion, the parameters used in the second-order expansion data shown here
are identical to those of Fig.~\ref{fig:benchmark} (see Tab.~\ref{tab:parameters_Kloss_1D}).}
\label{fig:taylor2}
\end{figure}%
In section~\ref{sec:neighbors}, it was mentioned that the order of the series expansion---call this $N_\text{series}$---%
inside of the effective Hilbert space $\mathcal{H}_\text{eff}(t)$ is much higher than the maximal neighbor order $m$
that is used in the construction of $\mathcal{H}_\text{eff}(t)$ as given in~\eqref{eq:H_eff}.
In practice, $m = 2$ is typically used for the recurring calculation of $\mathcal{H}_\text{eff}(t)$
(with $m_\text{init} \approx 10$ used for the construction of the initial $\mathcal{H}_\text{eff}(t = 0)$)
while $N_\text{series}$ is chosen dynamically based on a convergence criterion and commonly ranges from 20 to 60
(see Fig.~\ref{fig:error_sources}\textbf{c}).
Figure~\ref{fig:taylor2} demonstrates that not doing so drastically decreases the validity of the results,
i.e., if $N_\text{series}$ is chosen to be equal to $m = 2$ such that $U(\delta t)$ is approximated as
$\mathbbm{1} - \frac{i H \delta t}{\hbar} - \frac{1}{2}\left(\frac{H \delta t}{\hbar}\right)^2$,
then the results quickly turn unphysical.
This is primarily evidenced by the rapid growth of the wavefunction norm (see Fig.~\ref{fig:taylor2}\textbf{b}),
as such a low-order truncation no longer results in a unitary operator;
but even if the state is renormalized after each step,
the resulting state is completely incorrect, as demonstrated in Fig.~\ref{fig:taylor2}\textbf{a}.

Note that the second-order series-expansion results shown here still use the machinery of \mname{},
in particular the truncation and post-adaptation detruncation methods described in sec.~\ref{sec:detruncation}.
Even with a second-order method, state-vector truncation is necessary for time evolution beyond a handful of timesteps
due to the growth of the support of the state vector.
However, truncation becomes necessary after around eight timesteps, corresponding to $t \approx 0.06 \frac{2\pi}{\omega_0}$
for $\delta t = 0.05 / \omega_0$ down to $t \approx 0.012 \frac{2\pi}{\omega_0}$ for $\delta t = 0.01 / \omega_0$,
which is significantly earlier than the onset of the observed divergences.
This indicates that it is not the truncation, but rather the second-order expansion itself
that causes the unphysical behavior observed in Fig.~\ref{fig:taylor2}.

These results demonstrate that the essence of \mname{} is not captured by a low-order expansion of the time-evolution operator,
but lies instead in determining a useful reduced subspace that captures most of the state dynamics
and within which an essentially exact time evolution with $N_\text{Taylor} \gg m$
provides a highly accurate approximation of the exact dynamics that would take place in the underlying, unreduced Hilbert space,
as was motivated in sec.~\ref{sec:the_essence}.

\subsection{Series-based exponentiation compared to a Lanczos algorithm}\label{app:Lanczos}
As stated in App.~\ref{app:propagator}, the main allure of the computation
of $e^A v$ via repeated applications of $\frac{A}{n}$ to $v$ is that it is $\order{1}$ in memory,
independent of the order of the expansion.
Calculating the series expansion up to order $N_\text{series}$ is $\order{N_\text{series}}$ in time,
but due to the existence of highly efficient parallelized algorithms for this comparatively simple SpMV application,
the linear scaling in time has a small prefactor,
and instead memory is \emph{the} limiting factor for the entire method.
Therefore, the constant memory requirement outweighs the downsides of slightly higher cost in time.

Nonetheless, one could consider replacing the series-based evaluation of the single-step time-evolution operator $U(\delta t)$
described in App.~\ref{app:propagator} with a Lanczos-type decomposition~\cite{Noack,Paeckel2019}
\[ e^{-i\delta t H/\hbar}\ket{\psi(t)} \approx V_n e^{-i\delta t T_n/\hbar} V_n^\dag \ket{\psi(t)},\]
where $V_n$ is a $\left(\dim\{\mathcal{H}\} \times n\right)$-matrix that maps from the Krylov space $K^n$ to the Hilbert space,
and $V_n^\dag$ is its adjoint.
$T_n$ is the tridiagonal $\left(n\times n\right)$ matrix resulting from the Lanczos procedure,
similar to a partial diagonalization of $H$ within $K^n$.
The transformation into Krylov subspaces typically requires only relatively low dimensions on the order of 10,
smaller than the number of iterations we use for the Taylor expansion (see Fig.~\ref{fig:error_sources}\textbf{c}).
However, the Krylov method is computationally slightly more demanding than the series-based method:
The generation of $n+1$ Lanczos vectors also naturally requires $n$ applications of $H$ onto a vector,
as well as orthogonalization via the calculation of inner products and $\ell^2$ norms, which both require slightly more time.
The main issue, however, is that the construction of the $V_n$ matrix, which is composed of $n$ vectors in $\mathcal{H}_\text{eff}$,
is of order $\order{n}$ in both time \emph{and} memory, unlike the $\order{1}$ memory behavior of the series-expansion method.
This is ultimately the greatest benefit of the expansion method over the Lanczos method,
since on GPUs, highly parallelized applications of SpMV operations are cheap while memory is not.
For example, given an $\mathcal{H}_\text{eff}$ with a dimension of \num{7e7},
a single 128-bit complex-valued vector requires approximately \SI{1}{\gibi\byte} to store,
excluding the cost of the lookup table (App.~\ref{app:compression}).
If we were to construct a $V_n$ with a relatively modest $n=10$,
then this $V_n$ alone would consume over \SI{10}{\gibi\byte} in memory.
This may already constitute a significant portion of the total amount of available GPU memory,
which, on current devices, commonly ranges from \SIrange{8}{80}{\gibi\byte}---%
beyond that, a significant amount of RAM is still needed to store further matrices for observables, lookup tables, etc.
By contrast, the only additional memory requirements of the series expansion method of App.~\ref{app:propagator}
are that of a single temporary vector, i.e., \SI{1}{\gibi\byte} in this example.
However, one situation in which such a Krylov-type approach may prove advantageous even in terms of memory
is if an explicit construction of the Hamiltonian matrix within $\mathcal{H}_\text{eff}$ can be avoided entirely,
i.e., if the action of the underlying Hamiltonian can be evaluated exclusively in a matrix-free fashion
(and subsequently via the very small tridiagonal Lanczos matrix within the Krylov subspace).

Similarly, one could believe that the matrix $V_{n}$ should contain a smaller or identical number of non-zero elements as
the magnitude of $\bigcup_{k = 0}^{n-1} \mathcal{N}_{\ket{\psi(t)}}^{(k)}$ of~\eqref{eq:H_eff}
(or as the number of non-zero elements of the Hamiltonian on $\mathcal{H}_\text{eff}^{(n-1)}$),
and that, therefore, there should be no memory benefit in using~\eqref{eq:H_eff} rather than a Lanczos matrix.
However, because many of the states that make up $\mathcal{N}_{\ket{\psi(t)}}^{(k)}$
are already contained in $\mathcal{N}_{\ket{\psi(t)}}^{(k-1)}$, as not every neighbor state of a set is a new state---%
that is, because many typical neighbor states are in fact \emph{trivial} neighbor states---%
the magnitude $\abs{\bigcup_{k = 0}^m \mathcal{N}_{\ket{\psi(t)}}^{(k)}}$ is
typically far smaller than the sum of the magnitudes
$\sum_{k = 0}^m \abs{\mathcal{N}_{\ket{\psi(t)}}^{(k)}}$.
This is precisely the subadditive nature that was mentioned in section~\ref{sec:Krylov}
and which may cause the matrix $V_n$ to contain more non-zero elements than an
equivalent Hamiltonian on $\mathcal{H}_\text{eff}$.
For example, consider the following state with a TB Hamiltonian~\eqref{eq:tb} with zero on-site energy
\[\ket{\psi'} = \frac{1}{\sqrt{3}} \left[\ket{n-1} + \ket{n} + \ket{n+1}\right].\]
Note that $\ket{n \pm 1}$ are precisely the neighbor states of $\ket{n}$, meaning that the support of $\ket{\psi'}$
already contains trivial neighbors. This property then extends to the higher-order Krylov vectors:
The change-of-basis matrix $V_n$ for $n = 4$ is given by
\[V_4 = \frac{1}{12} \begin{pmatrix}
    0 &         0 & 0 & 8\\
    0 &         0 & 6 & 2\\
    0 &         3\sqrt{6} & 3 & -1\\
    4\sqrt{3} & -\sqrt{6} & 3 & -1\\
    4\sqrt{3} & 2\sqrt{6} & -6 & 2\\
    4\sqrt{3} & -\sqrt{6} & 3 & -1\\
    0 &         3\sqrt{6} & 3 & -1\\
    0 &         0 & 6 & 2\\
    0 &         0 & 0 & 8
    \end{pmatrix},
\]
where the rows correspond to the states $\{\ket{n - 4}, \ket{n - 3}, \ldots, \ket{n + 4}\}$.
This matrix is essentially dense, as 24 of the 36 matrix elements in $V_4$ are non-zero.
By contrast, the Hamiltonian in the equivalent $\mathcal{H}_\text{eff}^{(3)}$
has $d - 1$ non-zero elements in each of the first off-diagonals
(where $d = 9$ is the dimension of $\mathcal{H}_\text{eff}^{(3)}$), corresponding to 16 non-zero elements in total.
More generally, given this initial state, the dimension of $\mathcal{H}_\text{eff}^{(n-1)}$ will be $3 + 2(n-1)$
such that the number of non-zero elements of the Hamiltonian inside $\mathcal{H}_\text{eff}^{(n-1)}$
is $2(3 + 2(n-1) - 1) = 4n$; on the other hand,
each additional vector $v_k$ in the Lanczos decomposition appears to be comprised of $2k + 1$ non-zero elements,
meaning that the number of non-zero elements in the matrix $V_n$ is $n^2 + 2n$.
This means that the difference in density between $V_n$ and the Hamiltonian in $\mathcal{H}_\text{eff}^{(n-1)}$
continues to grow as $n$ is increased.

Nevertheless, two points should be noted about this behavior:
For one, it depends on the initial state, as an initial state such as $\frac{1}{\sqrt{2}} \left(\ket{n-100} + \ket{n+100}\right)$
would have resulted in a vastly different and much sparser $V_n$ matrix---though states that are delocalized among neighboring
or nearly neighboring states such as this $\ket{\psi'}$ could be considered quite typical states.
Second, if the underlying Hamiltonian itself is dense, then any explicit construction of the Hamiltonian inside
$\mathcal{H}_\text{eff}$ is likely also dense and thus almost certainly at a disadvantage relative to a Lanczos decomposition.
However, as discussed in detail in the main text, \mname{} relies on the Hamiltonian being sparse in various ways
and can generally be expected to fail for dense Hamiltonians;
moreover, as discussed in the introduction to App.~\ref{app:sparse},
the tight-binding Hamiltonian~\eqref{eq:tb} used for this example is a relatively dense Hamiltonian compared
to other typical many-body Hamiltonians, indicating that the sparseness advantage demonstrated above may,
in some situations, even be exceeded.

\begin{figure}
\centering
\includegraphics{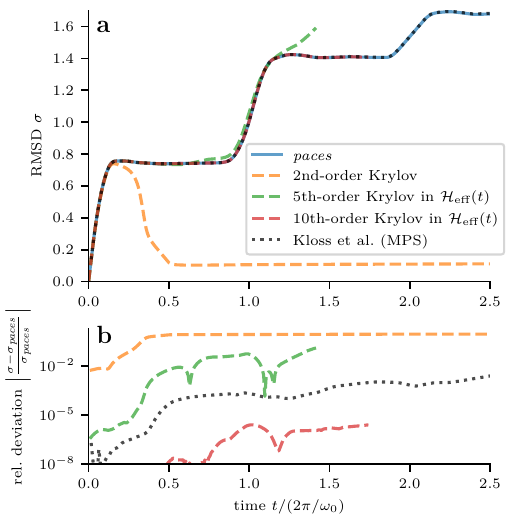}
\caption{%
The $g = 4\omega_0$ data shown in Fig.~\ref{fig:benchmark} produced using \mname{}
compared to data obtained using a Krylov expansion of $U(\delta t)$.
Both are computed within an effective Hilbert space with $m = 2$:
For second-order Krylov, this corresponds to a variant of a restarted Krylov procedure with truncation,
whereas the higher-order Krylov results can be understood as an alternative implementation
of the propagator $U(\delta t)$ within \mname{}.
The higher-order Krylov calculations terminated prematurely as a result of exceeding available memory.
\textbf{a}: RMSD results of the different approaches.
\textbf{b}: The relative deviations between the different results.
Multiset-MPS data of Kloss et al.~\cite{Kloss} is also shown for comparison.
Applicable parameters are identical to those of Fig.~\ref{fig:benchmark} (see Tab.~\ref{tab:parameters_Kloss_1D}).}
\label{fig:lanczos}
\end{figure}%
Finally, it should be pointed out that a very low-dimensional Krylov/Lanczos decomposition will typically be less dense
than the equivalent construction in $\mathcal{H}_\text{eff}$.
For the example given above, setting $n = 2$ causes $V_n$ to have only 8 non-zero elements (out of 10 in total),
which is identical to the number of non-zero matrix elements of $H$ in $\mathcal{H}_\text{eff}^{(1)}$,
negating any sparseness benefits.
However, as discussed in sec.~\ref{sec:Krylov}, an effective Hilbert space
$\mathcal{H}_\text{eff}^{(m)}$ of maximum order$m$ is often more powerful than the restriction to the Krylov space $K^{m+1}$:
To demonstrate this, the series-based exponentiation used by \mname{}
is explicitly compared to a Krylov-based alternative within $\mathcal{H}_\text{eff}^{(2)}$ in Fig.~\ref{fig:lanczos}.
The data shown there essentially showcases two distinct situations:
On the one hand, a second-order Krylov/Lanczos propagator
(by which we mean evolution in a three-dimensional subspace $\spn\{v, Hv, H^2 v\}$)
inside an effective space with $m = 2$ corresponds to a specific implementation of Krylov evolution
with periodically recomputed Krylov subspaces and in which the state vector between individual timesteps
is truncated and de-truncated as described in sec.~\ref{sec:detruncation}.
The higher-order Krylov methods shown in Fig.~\ref{fig:lanczos}, on the other hand, simply constitute an alternative version of \mname{}
in which the series-based computation of $U(\delta t)$ described in App.~\ref{app:propagator}
is replaced with a Krylov propagator that remains within the framework of \mname{}.
In the latter case, as the dimension of the Krylov space is increased,
the Krylov results should converge to the same result as those using the series-based computation of $U(\delta t)$---%
as is confirmed in Fig.~\ref{fig:lanczos}.

Similar to the second-order series results shown in App.~\ref{app:Taylor},
the second-order Krylov method reproduces the dynamics for a short amount of time
before starkly diverging from the correct results, underlining that an effective Hilbert space with $m = 2$
is capable of capturing much richer dynamics than a second-order Krylov method.
Furthermore, the fifth-order method is accurate over a longer time frame but still eventually diverges,
while the tenth-order Krylov results closely follow those obtained by \enquote{standard} \mname{}.
However, as was discussed above, the memory requirements of the Krylov method are higher than those of the series-based evolution,
meaning that the tenth-order Krylov method exceeded the available memory around $t = 1.75 \frac{2\pi}{\omega_0}$
whereas standard \mname{} using the same parameters was able to continue up to at least $8 \frac{2\pi}{\omega_0}$.
Note that the calculations shown here still constructed an explicit Hamiltonian matrix on $\mathcal{H}_\text{eff}^{(2)}$
and did not employ a \enquote{fully} matrix-free approach as mentioned above.
Such a combined matrix-free/Lanczos approach within the remaining framework of \mname{}
may prove beneficial in certain circumstances and would need to be studied in future work.

\section{How to weight states for truncation}\label{app:truncation}
The effective Hilbert space must be truncated at regular intervals to prevent it from exceeding available memory.
The simplest approach is to simply sort the expansion coefficients by their current weight
$w_0^{(n)} \coloneqq \abs{\braket{n}{\psi(t)}}^2$ and then truncate to the $M$ coefficients with the highest weight. 
Though this approach is certainly reasonable, it is not guaranteed to be optimal:
The basis states that exhibit the highest weight in the wavefunction at time $t$
are not necessarily those that bear the greatest importance for the future of the time evolution.
One might, therefore, consider estimating the future \enquote{importance} of the basis state $\ket{n}$
by weighting it not by $\abs{\braket{n}{\psi(t)}}^2$,
but rather by some method that assigns additional weight to highly connected states that may be explored in the next timestep,
for example:
\[
\abs{\braket{n}{\psi(t+\tau)}}^2 \approx \abs{\mel{n}{\left( 1 - \frac{i \tau}{\hbar} H \right)}{\psi(t)}}^2
    \eqqcolon w_\tau^{(n)}
.\]
The future-weighting parameter $\tau$ can be varied independently of the timestep $\delta t$, and
setting $\tau = 0$ results in the simple current-time weight $w_0^{(n)}$. 
However, in practice, said more general weighting mechanism performs best for $\tau = 0$%
---i.e., the naive approach appears to be superior to any \enquote{forward-looking} weighting.
Future analyses may uncover better weighting methods than the naive one, which would allow faster convergence to be achieved.

We point out a subtlety that must be taken into account when performing the truncation,
regardless of which weighting method is employed:
Especially in systems with high degrees of symmetry, the weighting method may result in degenerate weights,
and if there is a recurring but incidental order to the states prior to truncation,
the truncation may introduce unwanted biases.
For example, let the truncation happen after the $q$ highest-weighted basis states,
and let the $q$-th basis state be $\ket{n_1, \ldots, n_L}$.
If the weight of the basis state $\ket{n_1, \ldots, n_L + 1}$ is identical to that of $\ket{n_1, \ldots, n_L}$
and the states are always sorted lexicographically prior to the truncation,
then there will be an unwanted bias towards states with lower lexicographic order---in this case, the state $\ket{n_1, \ldots, n_L}$.
One way to circumvent this issue and avoid such a bias is to (pseudo\nobreakdash-)randomize
the order of degenerate-weight states around the cutoff point.

\section{Restriction errors due to finite neighbor orders}\label{app:restriction_error}
Using the terminology of sec.~\ref{sec:parallelization} and sec.~\ref{sec:error_analysis},
we provide a bound on the error induced by restricting the evolution of the state to
the restricted subspace $\mathcal{H}^{(m)}_\text{eff}(t)$ as given in~\eqref{eq:H_eff}.

This error can be quantified in the following way:
\[\varepsilon(P_m) \coloneqq \norm\Big{U(\delta t) \ket{\psi(t)} - \tilde{U}(\delta t) \ket{\psi(t)}}\]
where $\tilde{U}(\delta t)$ is the time-evolution operator restricted to the $m$-th-order subspace $\mathcal{H}^{(m)}_\text{eff}(t)$
and $P_m$ is the projector onto said subspace:
\[\tilde{U}(\delta t) \ket{\psi(t)}
    = \sum_{l=0}^\infty \frac{\left(-i \delta t/ \hbar\right)^l \left(P_m H P_m\right)^l}{l!}\ket{\psi(t)}.\]
By the construction of $\mathcal{H}^{(m)}_\text{eff}(t)$
(and neglecting the impact of the detruncation procedure of section~\ref{sec:detruncation}),
the projection acts trivially on all terms with indices $l \leq m$, such that
\[\left(P_m H P_m\right)^l \ket{\psi(t)} = H^l \ket{\psi(t)} \qfor{l \leq m},\]
and only for terms with $l > m$ may $\left(P_m H P_m\right)^l \ket{\psi(t)}$ and $H^l \ket{\psi(t)}$ differ.
Letting $Q_m \coloneqq \mathbbm{1} - P_m$, we can write
\begin{multline*}
\left(U(\delta t) - \tilde{U}(\delta t)\right) \ket{\psi(t)}\\
    = - \sum_{l=m+1}^\infty \sum_{\vec{\sigma}} \frac{\left(-i \delta t\right)^l A_{l-m}^{\vec{\sigma}}}{l! \hbar^l} H^m \ket{\psi(t)},
\end{multline*}
where
\[A_{l-m}^{\vec{\sigma}} \coloneqq (-Q_m)^{\sigma_1} H (-Q_m)^{\sigma_2} H \cdots (-Q_m)^{\sigma_{l-m}} H \]
and the sum over $\vec{\sigma}$ is over all strings $\left(\sigma_1, \sigma_2, \ldots, \sigma_{l-m}\right)$ of length $l-m$
with binary entries $\sigma_i \in \{0, 1\}$ where at least one of the $\sigma_i$ is non-zero.
We are now ready to determine an upper bound for the restriction error $\varepsilon(P_m)$:
Define
\begin{equation*}
u_l = \max_{\vec{\sigma}} \norm\big{A^{\vec{\sigma}}_{l-m} H^m \ket{\psi(t)}}.
\end{equation*}
If there exists some $\alpha \in \mathbb{R}^+$ such that $u_l \leq \alpha^l$ for any $l$, then the error is bounded by
\[\varepsilon(P_m)
    \leq \sum_{l=m+1}^\infty \frac{2^{l-m} - 1}{l!} \left(\frac{\alpha \delta t}{\hbar}\right)^{l},\]
or in the limit of small $\delta t$:
\[
\varepsilon(P_m) \leq \frac{1}{(m+1)!} \left(\frac{\alpha \delta t}{\hbar}\right)^{m+1},
\]
which is identical to~\eqref{eq:epsilon_Pm} in the main text.
The difficulty here will lie in determining $\alpha$. One may simplify the expression for $\alpha$
by further loosening the bound, since
\[u_l \leq \norm\big{H^l \ket{\psi(t)}} \leq \left(\norm{H}_\text{op}\right)^l\]
and therefore $\alpha \leq \norm{H}_\text{op}$, but then the bound is hardly useful in practical terms,
as it has lost all information about the underlying structure of the restricted Hilbert spaces.

\section{Finite lifetime via quantum jumps for nonradiative decay}\label{app:jumps}
To combat hard-cutoff artifacts in the spectra shown in section~\ref{sec:aggregates},
the computed correlation function $\expval{\mu(t) \mu(0)}$ is damped by an exponential function $e^{-t / \tau}$.
This procedure can, in principle, be justified as an application of the quantum-jump method~\cite{Plenio1998RMP,Breuer},
though with a slight caveat, as we shall demonstrate in this section:

Assume the overall dynamics of the nanoaggregate are given by the following zero-temperature Lindblad master equation
\[\pdv{\rho}{t}
    = -\frac{i}{\hbar} \comm{H}{\rho} + \gamma \sum_{j} \left[ L_k \rho {L_k}^\dag - \frac{1}{2} \acomm{{L_k}^\dag L_k}{\rho}\right]\]
with independent jump operators $L_k = \dyad{\Omega_\text{exc}}{k}$ that locally remove an exciton,
corresponding to uncorrelated local nonradiative decay of the chromophoric excitations.
Then the non-Hermitian contribution to the effective Hamiltonian in the quantum-jump procedure is given by
$-i\hbar \frac{\gamma}{2} \Pi_\text{1-exc}$, where $\Pi_\text{1-exc}$ is the projector onto the single-exciton manifold.
This additional operator simply decreases the norm of a single-exciton wavefunction with a lifetime $\tau = 2/\gamma$.
The effects of the actual quantum jumps, on the other hand, are irrelevant in this situation,
as the image of $L_k$ on any single-exciton state lies in the zero-exciton manifold,
i.e., $L_k \ket{\psi_\text{exc}} \otimes \ket{\psi_\text{vib}} = \ket{\Omega_\text{exc}} \otimes \ket{\psi_\text{vib}}$,
which does not contribute to the optical signal, since $\mel{\Omega_\text{exc}}{\mu}{\Omega_\text{exc}} = 0$.
Therefore, although a growing portion of the true state $\rho$ will in fact reside in the zero-exciton manifold,
we do not need to compute this part of the state for the optical signal, effectively circumventing the burdensome jump procedure.
To obtain sufficient damping over the duration of the ${\sim}\SI{460}{\fs}$ that are computed using \mname{},
we employ a decay rate of $\frac{1}{\tau} = 0.0577 \omega_0$ (equivalent to \SI{66.4}{\per\cm}),
which is well-separated from the system-dynamics timescales and leads to realistic linewidths in the final spectrum,
but corresponds to a nonradiative-decay-induced lifetime of only $\SI{80}{\fs}$,
which is roughly three orders of magnitude shorter than typical cyanine-dye exciton lifetimes.
Therefore, the lifetime imposed by this procedure must be understood as an effective broadening
that also incorporates other types of pathways beyond the non-radiative excitonic decay that it represents mechanistically.
Explicitly incorporating further decay channels such as spontaneous radiative decay
and damping of the vibrational modes via mechanisms such as intramolecular vibrational relaxation
would need to be the subject of future work.

\section{Further optical spectra and dynamics of H- and J-systems}
\subsection{Absorption characteristics of H- and J-dimers}\label{app:dimers}
To put the results presented in section~\ref{sec:aggregates} into perspective,
we recapitulate the $g$- and $J$-dependence of the absorption spectra of simple H- and J-coupled vibronic dimers.
These systems were the focus of much research in the 1960s~\cite{Merrifield1963RadRes,Kasha1963RadRes,Fulton1964JCP}
and have since served as the prototypical vibronic molecular aggregates~\cite{Schroeter2015PhysRep,Hestand2018ChemRev}.
We refer the reader to these references for a more comprehensive analysis of vibronic spectra in general.
Though vibronic dimers with a single oscillator per site by no means require methods such as \mname{}
to be solved on modern computing hardware, these simple systems serve as sanity checks for the method
and also provide background information to the reader who may be unfamiliar with the spectral properties of vibronic dimers.

All calculations presented in this appendix are computed using the methodology described in section~\ref{sec:aggregates}
with a simple two-site Holstein model.
Recall that the results presented in section~\ref{sec:aggregates} used $g = 0.71 \omega_0$ and $J = \pm 0.55\omega_0$,
which we also use as reference values in this section.
These parameters place those systems in the most interesting and complex intermediate-coupling regime in which neither the
vibronic nor the excitonic coupling dominates over the other.

\begin{figure}
\centering
\includegraphics{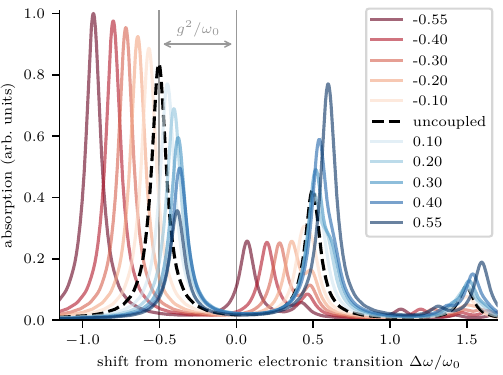}
\caption{The effect of the intermolecular excitonic coupling $J$ on two vibronic chromophores with $g = 0.71\omega_0$.
    The labels state $J$ as a fraction of the vibrational frequency $\omega_0$.}
\label{fig:Jseries}
\end{figure}
\begin{figure}
\centering
\includegraphics{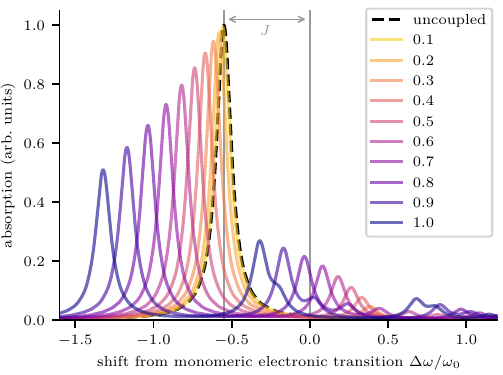}
\caption{The effect of the vibronic coupling $g$ on a J-type dimer with $J = -0.55\omega$.
    The labels state $g$ as a fraction of the vibrational frequency $\omega_0$.}
\label{fig:gseries_Jdimer}
\end{figure}
\begin{figure}
\centering
\includegraphics{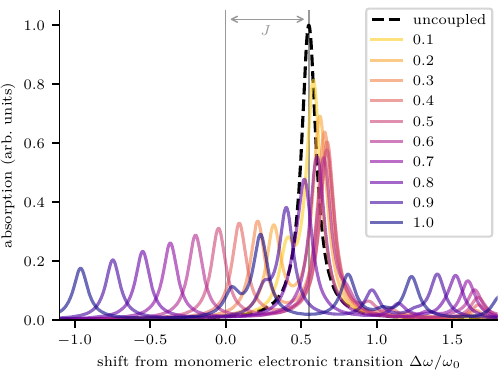}
\caption{The effect of the vibronic coupling $g$ on an H-type dimer with $J = 0.55\omega$.
    The labels state $g$ as a fraction of the vibrational frequency $\omega_0$.}
\label{fig:gseries_Hdimer}
\end{figure}

Figure~\ref{fig:Jseries} shows how the absorption spectrum changes from two electronically uncoupled monomers
to either an H-coupled (positive excitonic coupling $J > 0$) or J-coupled ($J < 0$) dimer.
The bare monomers show the expected Franck--Condon progression
with a regular spacing given by the vibrational (phonon) frequency $\omega_0$.
The zero-phonon line is red-shifted from the bare transition frequency by $g^2 / \omega_0$,
a quantity that is closely related to the Stokes shift~\cite{deJong2915PCCP}.
The primary effect of $J < 0$ is to red-shift the transitions in the familiar J-aggregate situation,
and vice versa for positive $J$, which gives rise to hypsochromic (H) shifts.
At the same time, J-coupling further concentrates intensity on the zero-phonon line
while H-coupling shifts intensity away from the zero-phonon line,
and increasing $\abs{J}$ in either direction causes the all but the zero-phonon line to split into multiplets.

Approaching the intermediate-coupling regime from the other side, Figures~\ref{fig:gseries_Jdimer} and \ref{fig:gseries_Hdimer}
show the effect of increasing vibronic coupling on a J- and H-coupled dimer, respectively.
In the absence of vibrations, both spectra show a single absorption line that is shifted from the
bare monomeric transition frequency by exactly $J$.
For a J-dimer, the bright state $\left(\ket{0} + \ket{1}\right) / \sqrt{2} \propto \mu \ket{\Omega_\text{exc}}$
represents the lower of the two single-exciton eigenstates, hence the red-shifted absorption line,
while the bright state in the H-dimer is given by the higher-energy eigenstate.
Adding vibronic coupling to the J-dimer (Fig.~\ref{fig:gseries_Jdimer}) further red-shifts the main absorption lines
and also leads to the gradual formation of 0-1, 0-2, etc.~transitions, which are themselves split due to the excitonic coupling.
On the other hand, the effect of vibronic coupling on the H-dimer (Fig.~\ref{fig:gseries_Hdimer}) is more complex:
The main absorption line is asymmetrically split into two lines, with the higher-energy line first becoming slightly blue-shifted
for small $g$ before also red-shifting for large $g$; the lower-energy line of said pair is consistently red-shifted.
At the same time, higher-phonon transition lines are formed and cross into the same frequency range as the uncoupled transition.

\subsection{Invariance of localized-state dynamics}\label{app:video}
\begin{figure}
\centering
\includegraphics{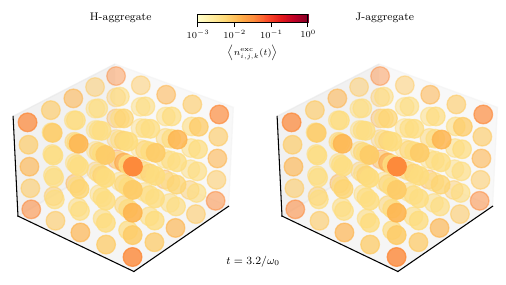}
\caption{(Multimedia available online/see the ancillary file \texttt{both\_agg\_5x5x5.mp4})
The evolution of the  exciton distribution $\expval{n_{i,j,k}^\text{exc}(t)}$
resulting from an initially fully localized Franck--Condon excitation in H- or J-type $5 \times 5 \times 5$ cubic lattices.
Parameters are identical to those used for the symmetrically delocalized initial state in the $5 \times 5 \times 5$ lattice
(see Tab.~\ref{tab:parameters_abs_spec_3d} in App.~\ref{app:parameters}).}
\label{fig:both_agg_n5x5x5}
\end{figure}

In Fig.~\ref{fig:both_agg_n5x5x5} (Multimedia available online),
the exciton dynamics resulting from an initially fully localized Franck--Condon excitation
$\ket{j_\text{cent}} \otimes \ket{\Omega_\text{vib}}$,
where $j_\text{cent}$ is the index of the central site of a $5 \times 5 \times 5$ lattice, are visualized.
As was established in sec.~\ref{sec:aggregates}, the H- and J-aggregate dynamics are completely identical
as long as the initial state is fully localized, extending previous 1D results~\cite{Kessing2022JPCL}.

\subsection{Geometric effects of eight-site aggregates}\label{app:n8}
\begin{figure}
\centering
\includegraphics{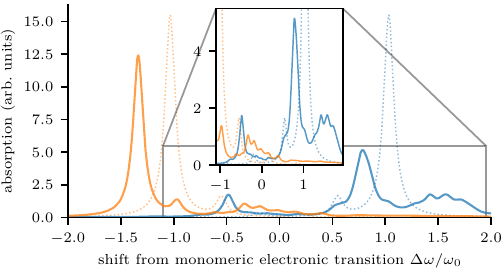}
\caption{
The optical spectrum of a 1D chain with eight sites.
Apart from the arrangement of the sites into a chain rather than a cube,
this corresponds exactly to the system shown in Fig.~\ref{fig:abs_spec_4}\textbf{a}.
Blue lines represent H-aggregate, orange lines J-aggregate; solid lines are vibronic compounds and dotted lines purely electronic.
Full list of parameters is given in App.~\ref{app:parameters}.}
\label{fig:abs_spec_n8}
\end{figure}

In section~\ref{sec:aggregates}, we found that the $2 \times 2 \times 2$ H-aggregate leads to a
rich optical spectrum (shown in Fig.~\ref{fig:abs_spec_4}\textbf{a}).
In Fig.~\ref{fig:abs_spec_n8}, we show the optical spectrum of an equivalent eight-site system that has been
unraveled into a one-dimensional chain.
The spectrum of this 1D system differs markedly from that of the $2 \times 2 \times 2$ system
and is much more reminiscent of the dimers shown in App.~\ref{app:dimers},
confirming that the rich structure of its 3D equivalent is, in fact, due to the geometry.
However, one might note that the J-aggregate spectrum of the eight-site chain presents some additional features around
$\Delta \omega = 0$ which are not present in the $2 \times 2 \times 2$ case.


\section{Parameters \& runtimes}\label{app:parameters}
In this appendix, we state the parameters used to produce the data shown in the various Figures used throughout
this publication.
Throughout this Appendix:
\begin{itemize}
    \item $N$ is the number of sites (chromophores) in the system.
    \item $d_\text{pho}$ is the hard cutoff dimension of the vibrational sites
        (i.e., $d_\text{pho} - 1$ is the maximum number of phonons at each site).
    \item $g$, $J$, and $\omega_0$ are the vibrational-coupling, excitonic-coupling and vibrational angular frequency, respectively,
        as given in~\eqref{eq:holstein}.
    \item $m_\text{init}$ is the maximum neighbor order used for the initial creation of the effective Hilbert space
        prior to the first timestep in the calculation (see~\eqref{eq:H_eff}).
    \item $m$ is the maximum neighbor order used recurrently at every timestep in the calculation.
    \item \qnom{} is the nominal truncation number (see~\eqref{eq:qtrue}) prior to expanding
        the effective Hilbert space via~\eqref{eq:H_eff}.
    \item $\delta t$ is the length of a timestep in the calculation.
    \item $t_\text{max}$ is the maximum simulation time, stated in units of ${\omega_0}^{-1}$,
        and $T_R$ is the total elapsed real time (the runtime in the practical sense) that \mname{}
        needed to reach $t_\text{max}$ when running on an Nvidia A100 (\SI{80}{\gibi\byte} HBM2e).
\end{itemize}
For some system/parameter combinations, the \mname{} calculation would exceed the available GPU RAM.
In such cases, the calculation can often still be performed by using \emph{unified memory},
which shares GPU and CPU RAM, providing a significant increase in effective available memory
at the cost of runtime due to a considerable rise in memory transfer time.
If a certain calculation required the use of unified memory, it will be stated explicitly;
otherwise, it ran solely on GPU RAM.

\begin{table}
\caption{The parameters used in the 1D calculations shown in Fig.~\ref{fig:benchmark}
(some of which are also shown in Figs.~\ref{fig:coeff_distribution}, \ref{fig:error_sources}, \ref{fig:convergence_q},
\ref{fig:taylor2}, \ref{fig:lanczos} and \ref{fig:conv_n25_d128_g4}).
$\chi'$ is the multiset MPS bond dimension used by Kloss et al.~\cite{Kloss} in their calculations
that were used as a benchmark.
The following were constant for all of these calculations:
$J = \omega_0$, $m_\text{init} = 10$, $m = 2$, $\delta t = 0.05/\omega_0$
(Kloss et.~al~\cite{Kloss} use $\delta t = 0.1/\omega_0$).
The runtime $T_R$ is the time \mname{} required to evolve up to $t_\text{max} = 50/\omega_0$---%
note that Fig.~\ref{fig:benchmark} only shows data up to $t = 40/\omega_0$.}
\begin{center}
\setlength{\tabcolsep}{6pt}
\begin{tabular}{l r r r r r r r}
\toprule 
$g/\omega_0$&   1.0 & 1.5 & 2.0 & 2.5 & 3.0 & 3.5 & 4.0\\ \midrule
$d_\text{pho}$& 16  & 16  & 64  & 64  & 64  & 64  & 64\\
$N$   &         75  & 75  & 51  & 25  & 25  & 25  & 25\\
\qnom{}/$10^6$& 25  & 25  & 25  & 32  & 32  & 32  & 32\\
$T_R$/minutes & 84  & 85  & 87  & 94  & 98  & 109 & 117\\
$\chi'$ of ref.~\onlinecite{Kloss}
              & 16  & 16  & 16  & 32  & 32  & 32  & 32\\
\bottomrule
\end{tabular}
\end{center}
\label{tab:parameters_Kloss_1D}
\end{table}
Tab.~\ref{tab:parameters_Kloss_1D} lists the parameters used in the computation of the 1D Holstein systems
that are primarily shown in Fig.~\ref{fig:benchmark} of the main text.

The dimer reference calculations shown in App.~\ref{app:dimers} use
$q_\text{nom} = \num{16000000}$, $m_\text{init} = 6$, $m = 2$, $d_\text{pho} = 16$, $\delta t = 0.05/\omega_0$ and
$N = 2$ throughout. The excitonic and vibronic coupling $J$ and $g$ are stated in the Figure captions.

The following parameters were used for all of the calculations of $\expval{\mu(t) \mu(0)}$ of the 1D and 2D systems
shown in Fig.~\ref{fig:abs_spec_3}:
$g = 0.71\omega_0$, $d_\text{pho} = 16$, $\delta t = 0.05/\omega_0$,
$m_\text{init} = 10$, $m = 2$, $q_\text{nom} = 16 \times 10^6$,
and the total number of sites was $N = 64$.
$J = 0.55\omega_0$ was used for the H-aggregates and $J = -0.55\omega_0$ for the J-aggregates.
The runtime $T_R$ that \mname{} required to reach $t_\text{max} = 100/\omega_0$ was
107 (105) minutes for the 1D H-aggregate (J-aggregate) and
214 (211) minutes for the 2D H-aggregate (J-aggregate).
The 1D system of 8 sites shown in Fig.~\ref{fig:abs_spec_n8} used identical parameters apart from the number of sites.
Its runtime was 33 minutes for both the H and J aggregate.

The parameters for the 3D $4 \times 4 \times 4$ system that was shown in Fig.~\ref{fig:abs_spec_3}
as well as the remaining data shown in Fig.~\ref{fig:abs_spec_4} are given in Tab.~\ref{tab:parameters_abs_spec_3d}.

\begin{table}
\caption{The runtime $T_R$ that \mname{} required to reach $t_\text{max} = 100/\omega_0$ in the 
calculations underlying the optical spectra of 3D systems shown in Fig.~\ref{fig:abs_spec_4},
and whether or not unified memory was required.
$J = 0.55\omega_0$ was used for the H-aggregates and $J = -0.55\omega_0$ for the J-aggregates.
Apart from that, the parameters were constant for all of these calculations:
$g = 0.71\omega_0$, $d_\text{pho} = 16$, $\delta t = 0.05/\omega_0$,
$m_\text{init} = 6$, $m = 2$, $q_\text{nom} = 16 \times 10^6$.}
\begin{center}
\setlength{\tabcolsep}{3.9pt}
\begin{tabular}{l cc cc cc cc}
\toprule 
$N$ &   \multicolumn{2}{c}{$2\times 2 \times 2$} & \multicolumn{2}{c}{$3\times 3 \times 3$}
        & \multicolumn{2}{c}{$4\times 4 \times 4$} & \multicolumn{2}{c}{$5\times 5 \times 5$}\\ \midrule
                & H  & J  & H   & J   & H   & J   & H   & J\\
$T_R$/minutes   & 42 & 42 & 149 & 149 & 539 & 523 & 2048 & 2021\\
unified mem.  & no & no & no  & no  & yes & yes & yes  & yes\\
\bottomrule
\end{tabular}
\end{center}
\label{tab:parameters_abs_spec_3d}
\end{table}

\section{Convergence}\label{app:convergence}
\subsection{Convergence with respect to the neighbor order \texorpdfstring{$m$}{m}}
\begin{figure}
\centering
\includegraphics{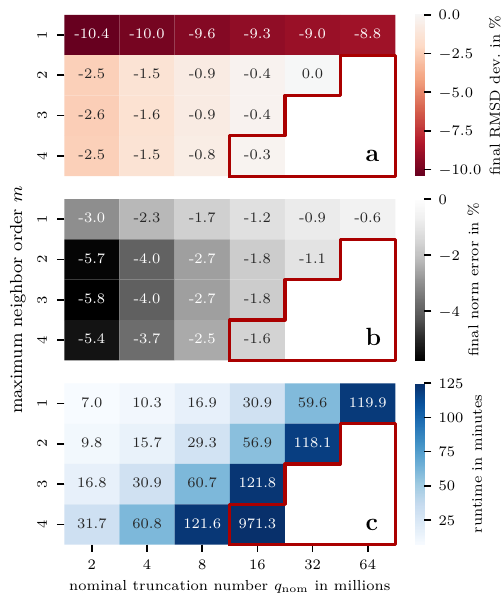}
\caption{Relative RMSD deviations (\textbf{a}), change in norm (\textbf{b}) and runtime (\textbf{c})
or the $g = 4$ calculation shown in Fig.~\ref{fig:benchmark}
as a function of convergence parameters. Data refers to $t = 50/\omega_0$.
The RMSD deviation in~\textbf{a} is measured with respect to the $m = 2$, $q_\text{nom} = 32 \times 10^6$ data.
The red lines indicate the region that exceeds the available GPU RAM,
and the $m = 4$, $q_\text{nom} = 16 \times 10^6$ calculation was obtained using unified memory,
which is also the reason for its drastically increased runtime.
Note that the initial maximum neighbor order at initialization prior to the first timestep of the calculation is chosen as
$m_\text{init} = 12 - m$ here, such that each calculation had the same $\mathcal{H}_\text{eff}$ during its first timestep.}
\label{fig:conv_n25_d128_g4}
\end{figure}
\begin{figure*}[t]
\centering
\includegraphics{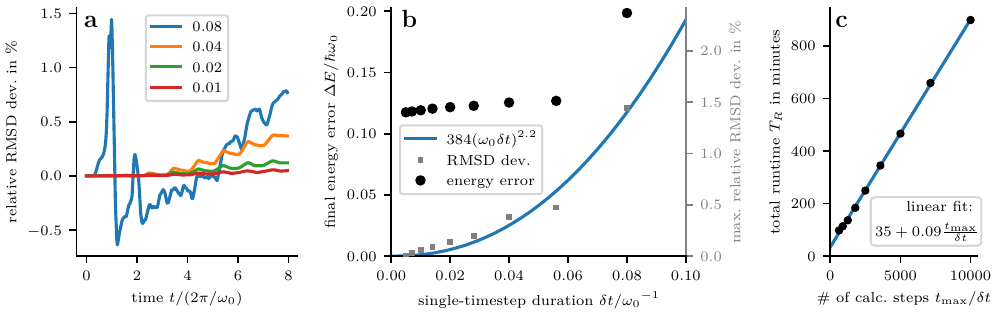}
\caption{Convergence as a function of the single-timestep duration $\delta t$.
Data shown in \textbf{b} and \textbf{c} is for $t = 50/\omega_0$.
RMSD deviations in this figure are relative to the data produced using the smallest timestep of $\delta t = 0.005/\omega_0$.
Apart from $\delta t$, this calculation is identical to the $g = 4\omega_0$ calculation of Fig.~\ref{fig:benchmark}
(parameters given in Tab.~\ref{tab:parameters_Kloss_1D}).
\textbf{a}: The deviations in RMSD over time for selected values of $\delta t$,
indicated in the legend in units of $1/\omega_0$.
\textbf{b}: The final error in the total energy (black circles) and the maximum relative deviation in RMSD (gray squares),
along with a monomial least-squares fit.
\textbf{c}: The runtime (on an Nvidia A100) as a function of the number of timesteps
(proportional to $1/\delta t$) along with a linear least-squares fit.}
\label{fig:convergence_dt}
\end{figure*}%
Figure~\ref{fig:conv_n25_d128_g4} shows the convergence and runtime of the $g = 4\omega_0$ calculation shown in
Figs.~\ref{fig:coeff_distribution}, \ref{fig:benchmark}, \ref{fig:error_sources} and \ref{fig:convergence_q}
as a function of the truncation parameter \qnom{} and the maximum neighbor order $m$.
Choosing $m = 1$ leads to markedly different RMSD results while all $m \geq 2$ yield essentially constant results with respect to $m$.
The convergence as a function of \qnom{} scales identically for all $m \geq 2$,
but the runtime approximately doubles with every increment of $m$, indicating that $m = 2$ is the optimal choice for this system.
Interestingly, the norm seems to be better-conserved for $m = 1$ than for the other calculations:
This is likely evidence of the \textit{good convergence to an incorrect result} that was mentioned in section~\ref{sec:error_analysis}.
Note that, as was pointed out in sec.~\ref{sec:neighbors}, the use of $m = 2$ is not equivalent
to a second-order Taylor expansion of the time-evolution operator $U(\delta t)$ (see App.~\ref{app:Taylor}),
nor is it equivalent to a second-order Krylov expansion (see App.~\ref{app:Lanczos}).

\subsection{Convergence with respect to the timestep duration \texorpdfstring{$\delta t$}{dt}}
Figure~\ref{fig:convergence_dt} demonstrates the convergence of the calculation as a function of the single-timestep duration $\delta t$.
In practice, it turns out that $\delta t$ has little effect on the calculation as long as a certain threshold is reached
(compare $\Delta E$ for $\delta t < 0.06/\omega_0$ to the $\delta t = 0.08/\omega_0$ datapoint in Fig.~\ref{fig:convergence_dt}),
and, if $\delta t$ is then further increased beyond a second threshold, the single-timestep evolution as described in
App.~\ref{app:propagator} simply fails to converge correctly, causing the norm of the state to explode---%
for the system shown in Fig.~\ref{fig:convergence_dt}, this happened around $\delta t \approx 0.16 / \omega_0$.

\subsection{Optical spectra} \label{app:conv_spectra}
\begin{figure}
\centering
\includegraphics{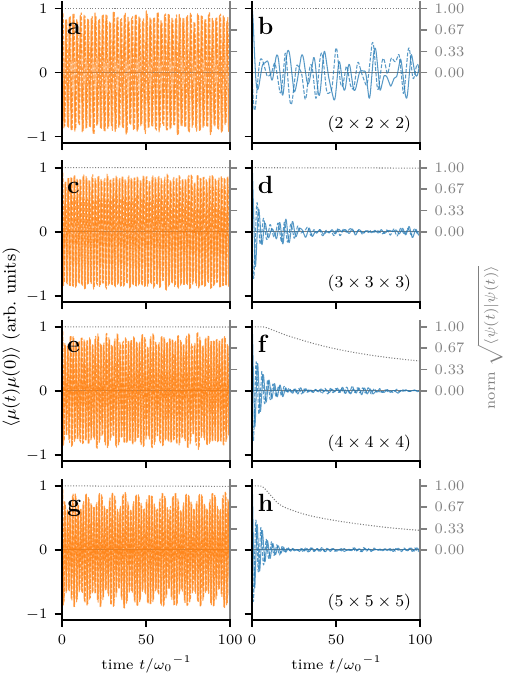}
\caption{The real-time computed signals $\expval{\mu(t) \mu(0)}$ used in Fig.~\ref{fig:abs_spec_4}.
Left-hand side (\textbf{a}, \textbf{c}, \textbf{e}, \textbf{g}) shows J-aggregates;
right-hand side (\textbf{b}, \textbf{d}, \textbf{f}, \textbf{h}) shows H-aggregates.
Solid lines are the real part of the signal and dashed lines are the imaginary part.
Dotted gray lines are the norm of the wavefunction.
Note that the signal $\expval{\mu(t) \mu(0)}$ has been renormalized,
that is, the decay in the signal is \emph{not} due to the decaying norm.}
\label{fig:rt_signal_4}
\end{figure}

Figure~\ref{fig:rt_signal_4} shows computed real-time optical responses $\expval{\mu(t) \mu(0)}$;
this is the data that was used for Fig.~\ref{fig:abs_spec_4} before being damped by $e^{- t/\tau}$ and Fourier-transformed.
The differences in the true lifetime between the H- and J-aggregate signals is striking,
as is the difference in the decay of the norm,
especially between Fig.~\ref{fig:rt_signal_4}\textbf{g} and \textbf{h}.
This indicates that \mname{} struggled with the H-aggregate much more than with the equivalent H-aggregate,
which is surprising at first glance given that the two Hamiltonians have exactly the same structure and connectivity
and differ only in the sign of one of the couplings.
The difference is likely due to the following:
In a purely electronic J-aggregate, the initial state after the optical excitation, $\sum_j \ket{j} \otimes \ket{\Omega_\text{vib}}$
overlaps most significantly with the lowest-energy state of the excitonic aggregate.
Thus, even in the presence of weak to moderate vibronic coupling, most other states in the Hilbert space
have higher energy than the initial state and are, therefore, never significantly explored by the state,
meaning that it mostly stays localized on basis states close to where it already is.
By contrast, in an equivalent H-aggregate, the state after the initial excitation is at the top of the excitonic band.
Therefore, the state can essentially trade excitonic energy for vibrational energy and explore a much larger
fraction of the Hilbert space.

\begin{figure}
\centering
\includegraphics{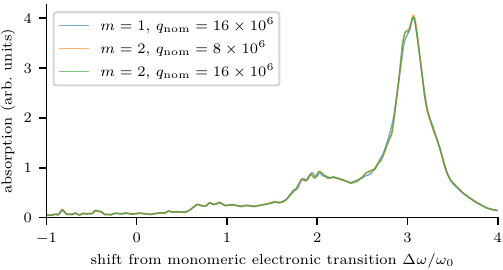}
\caption{Spectra of the $5 \times 5 \times 5$ H-aggregate shown in Fig.~\ref{fig:abs_spec_4}\textbf{d}
for different convergence paramters.
The green line corresponds to the $m$ and \qnom{} shown in the main text
while the blue (orange) line shows the effect of halving $m$ (\qnom{}).
The lifetime $\tau$ was increased to \SI{160}{\fs} because
$\tau = \SI{80}{\fs}$ as in the main text would render differences essentially invisible.
This is the least well-converged of all optical spectra shown in sec.~\ref{sec:aggregates}.}
\label{fig:conv_n5x5x5_abs_spec}
\end{figure}

However, as shown in Fig.~\ref{fig:rt_signal_4}\textbf{f} and \textbf{h}, though the norm of the state decays quite unfavorably,
so does the signal itself. As a result, the loss in fidelity has little impact on the measurable of interest.
We verify this by investigating the convergence of the Fourier-transformed optical spectra of the most unfavorable case,
the $5 \times 5 \times 5$ H-aggregate, in Fig.~\ref{fig:conv_n5x5x5_abs_spec},
which demonstrates that, despite the significant loss in norm over time, the spectrum itself is still sufficiently well-converged.

\end{document}